\documentclass[aps,prl,reprint,groupedaddress,twocolumn,showkeys,amsmath,amssymb,longbibliography,floatfix]{revtex4-1}
\usepackage[T1]{fontenc}
\usepackage{graphicx}
\usepackage{dcolumn}
\usepackage{amsmath,amssymb,amsfonts,bm,dsfont,units}
\usepackage{hyperref}
\hypersetup{colorlinks=true,citecolor=blue,linkcolor=blue,urlcolor=blue,pdfstartview=FitH}
\usepackage{appendix} 
\usepackage{braket}
\usepackage{float,latexsym}
\usepackage{multirow}
\usepackage[normalem]{ulem} 
\usepackage[caption=false]{subfig} 
\usepackage{xcolor}
\newcommand{\vect}[1]{\boldsymbol{#1}}
\usepackage{array}
\usepackage{graphicx} 
\usepackage{hhline,booktabs}
\usepackage{makecell}
\usepackage{wasysym}

\definecolor{darkred}{rgb}{0.8, 0.0, 0.0}

\newcommand{\refcite}[1]{Ref.~\cite{#1}}
\newcommand{\refscite}[1]{Refs.~\cite{#1}}

\newcommand{\figref}[1]{Fig.~\ref{#1}}
\newcommand{\sfigure}[2]{Figure~\hyperref[#1]{\ref{#1}(#2)}}
\newcommand{\sfigref}[2]{Fig.~\hyperref[#1]{\ref{#1}(#2)}}

\usepackage{color}
\definecolor{dkgreen}{rgb}{0,0.5,0}
\definecolor{midnightblue}{rgb}{0.39,0.58,0.93}

\definecolor{kspink}{RGB}{200,0,200}
\definecolor{appendixgreen}{RGB}{9, 145, 58}

\newcommand{\comment}[1]{}{}
\newcommand{\vacflux}{\kern-0.2em$\varhexagon$\kern-0.36em\raise1.07ex\hbox{$\varhexagon$}\kern-0.36em\hbox{$\varhexagon$}\kern-0.77em\raise0.14ex\hbox{\textcolor{white}{$\bullet$}}\kern-0.47em\raise0.43ex\hbox{\textcolor{white}{$\bullet$}}\kern-0.61em\raise0.43ex\hbox{\textcolor{white}{$\bullet$}}}
\begin{document}

\title{Vacancies in generic Kitaev spin liquids}

\author{Ihor Yatsuta}
\affiliation{Department of Condensed Matter Physics, Weizmann Institute of Science, Rehovot 7610001, Israel}
 \author{David F. Mross}
\affiliation{Department of Condensed Matter Physics, Weizmann Institute of Science, Rehovot 7610001, Israel}

\date{\today}

	
	\begin{abstract}
		The Kitaev honeycomb model supports gapless and gapped quantum spin liquid phases. Its exact solvability relies on extensively many locally conserved quantities. Any real-world manifestation of these phases would include imperfections in the form of disorder and interactions that break integrability. We show that the latter qualitatively alters the properties of vacancies in the gapless Kitaev spin liquid: (i) Isolated vacancies carry a magnetic moment, which is absent in the exactly solvable case. (ii) Pairs of vacancies on even/opposite sublattices gap each other with distinct power laws that reveal the presence of emergent gauge flux.
	\end{abstract}
	\maketitle

{\bf Introduction.} Quantum spin liquid (QSL) behavior has been reported for an increasing number of materials \cite{Balents2010,Savary2016,Norman2016,Zhou2017,Takagi2019,Knolle2019,Broholm2020}.  
These phases are characterized not by their broken symmetries but rather by their fractional quasiparticles and the associated topological properties. QSLs with a bulk gap are sharply distinct from conventional phases by their long-range entanglement. However, this criterion is impractical for experimental identifications, and positively identifying QSLs remains a formidable challenge. 

 More direct evidence of QSL could arise from gapless (bulk) excitations that respond to weak experimental probes. 
On the flip side, gapless bulk excitations pose significant challenges for theoretical descriptions. The understanding is primarily derived from poorly-controlled field-theoretical treatments or rather special exactly-solvable models. The best-known example of the latter is the Kitaev honeycomb model \cite{Kitaev2006}, whose exact solution relies on its many conserved quantities. 

Pioneering work has argued that the highly anisotropic spin-spin interactions of the honeycomb model may arise in strongly spin-orbit coupled Mott insulators. In particular, the honeycomb model is proposed to approximate the local environment experienced by effective spin-1/2 moments in various iridates and $\alpha$-RuCl$_3$ (see \refscite{Banerjee2016,Winter2017,Hermanns2018,Takagi2019,Motome2020,Trebst2022,Kasahara2018,Kitagava2018,Takahashi2019,Yamada2020,Do2020,Yokoi2021,Bruin2022,Czajka2021,Czajka2023}). The prospect of realizing a Kitaev QSL demands a careful analysis of which properties of the finely-tuned toy model are generic and, therefore, constitute predictions for actual experiments. Real materials will inevitably deviate from the honeycomb model in at least two ways \cite{PhysRevLett.105.027204,PhysRevLett.112.077204,Jiang2011,Tikhonov2011,Chaloupka2013,Rau2014,Knolle2018,Wang2019,Gordon2019,Hickey2019,Hwang2022}. Firstly, the many conservation laws are violated, and secondly, there will be disorder.

The implications of the former were analyzed in \refscite{Chaloupka2013,Song2016,Gotfryd2017,Gohlke2017,Zhang2021}, which showed that generic perturbations qualitatively change key observables without destabilizing the phase. In particular, the honeycomb model exhibits a hard spin gap and ultra-short range spin-spin correlations. By contrast, the spin gap is absent in a generic incarnation of the phase, and correlations decay as power laws \cite{Song2016}.

Concurrently, disorder effects were extensively studied for the Kitaev honeycomb model \refscite{Willans2010,Willans2011,Sreejith2016,Nasu2020,Kao2021,Singhania2023,Zschocke20152}. An important role is played by vacancies, which dominate the thermodynamic properties at low temperatures and weak magnetic fields. In the gapless phase, an isolated vacancy yields an exact two-fold ground state degeneracy protected by Kramers' theorem but no moment at zero magnetic field. A weak field induces a moment with a singular field dependence. Moreover, a finite number of vacancies leads to a large number of exact zero-energy states. Ref.~\onlinecite{Dhochak2010} constructed a full set of separately conserved Pauli operators for each vacancy, which implies an extensive ground state degeneracy for a finite vacancy concentration. Such a situation is thermodynamically unstable and cannot withstand generic perturbations.

Our work demonstrates that a generic gapless Kitaev QSL with vacancies displays qualitatively different behavior than the fine-tuned honeycomb model. Most importantly, we show that vacancies exhibit a non-zero magnetic moment at zero field. Such a moment is allowed on symmetry grounds. Indeed, vacancies in the gapped phase of the honeycomb model, which has the same symmetries, trap a non-zero moment. The moment's magnitude $M$ is of order unity in the zero-correlation length limit and decreases with $\xi$ \cite{Willans2010,Willans2011}. We note that in systems with a finite correlation length, such as valence bond solids, even weak bond disorder may be sufficient to induce magnetic moments \cite{Kimchi2018}.

By contrast, in gapless QSLs, the fate of such a moment is unclear. The localization length diverges, and in the integrable honeycomb model, $M(\xi \rightarrow \infty)=0$. We show that generic perturbations yield a non-zero moment despite the diverging length. Moreover, our work shows that vacancy-induced magnetization falls off according to a universal power law that depends on whether the vacancy traps an emergent $\mathbb{Z}_2$ gauge flux. Finally, multiple vacancies interact with an RKKY-like interaction characterized by flux-dependent power laws, which lifts the unphysically large ground-state degeneracy of the honeycomb model with vacancies.

\begin{figure}[htb]
   \centering \includegraphics[width=0.99\linewidth]{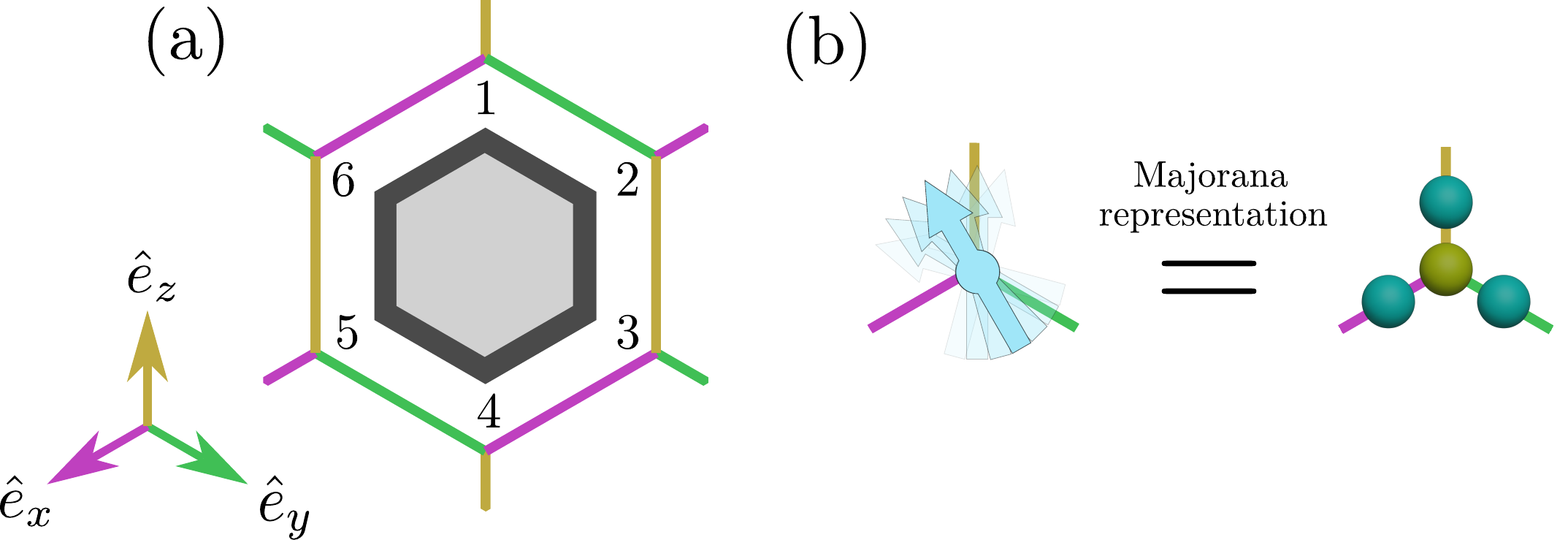}
   \caption{(a) The honeycomb lattice features three nearest-neighbor link directions, which we denote by $\hat e_{x}, \hat e_{y},\hat e_{z}$ as indicated. The operators $
\hat W_\text{\varhexagon} \equiv \sigma^z_{1}\sigma^x_2\sigma^y_3\sigma^z_4\sigma^x_5\sigma^y_6$ for any hexagonal plaquette commute with the Kitaev Hamiltonian of Eq.~\eqref{hkitaev}. (b) To diagonalize $H_\text{K}$, each spin is represented by four Majorana fermions, subject to a local constraint. It is suggestive to associate three `gauge Majoranas' with the link directions and the remaining `matter Majorana' with the site center.}
   \label{fig: plaquette_operator}
 \end{figure}

{\bf Kitaev honeycomb model.}
 To begin, we briefly recall the celebrated Kitaev honeycomb model and its solution. The Hamiltonian is
\begin{align}
H_\text{K} =J \sum_{ \vect r \vect r'}  \sigma^{\mu}_{\vect r} \sigma^{\mu}_{\vect r'}~\qquad \vect r' = \vect r + \hat e_\mu ,
\label{hkitaev}
\end{align}
where $\hat e_\mu$ with $\mu=x,y,z$ represent the three link directions on a honeycomb lattice. The model $H_\text{K}$ exhibits one conserved $\mathbb{Z}_2$ `flux' $
\hat W_\text{\varhexagon} $
for any unit cell (see \figref{fig: plaquette_operator}).
To express the remaining degrees of freedom, we encode each Pauli operator in four Majorana fermions $
\vec \sigma_{\boldsymbol{r}} = i \lambda_{\boldsymbol{r}}\vec \xi_{\boldsymbol{r}}$ with the local constraint ${\cal D}_{\boldsymbol{r}}\equiv\lambda_{\boldsymbol{r}}
\xi_{\boldsymbol{r}}^x
\xi_{\boldsymbol{r}}^y
\xi_{\boldsymbol{r}}^z=1 $. The Hamiltonian takes the free-fermion form 
\begin{align}
H_\text{K} =-J \sum_{ \vect r \vect r'} i  \lambda_{\vect r} \hat u_{\vect r \vect r'}\lambda_{\vect r'} \qquad   \hat u_{ \vect r \vect r'}\equiv i \xi^{\mu}_{\vect r} \xi^{\mu}_{\vect r'}~.
\label{eq.free}
\end{align}
The link variables $\hat u_{\vect r \vect r'}=\pm 1$ commute with the Hamiltonian but \textit{anticommute} with the constraints on sites ${\boldsymbol{r}}$ and ${\boldsymbol{r}'}$. The fermions $\lambda_{\vect r}$ likewise anticommute with ${\cal D}_{\boldsymbol{r}}$. Consequently, Eq.~\eqref{eq.free} describes a $\mathbb{Z}_2$ gauge theory where the gauge-invariant flux $\hat W_\text{\varhexagon}$ is static. In the ground state, $\hat W_\text{\varhexagon}=1$ on all plaquettes, and the $\lambda$ fermions form two Majorana cones at the Brillouin zone corners $\vect{K},\vect{K}'$. 

{\bf Generic Kitaev QSLs.} The gapless Kitaev QSL is stable against any weak time-reversal invariant perturbations to $H_\text{K}$. It is useful to distinguish between three kinds of such modifications.

\textit{1. Additional free fermions terms.} Multi-spin interactions such as \begin{equation}\delta H_{2} \sim \sigma^y_1 \sigma^{x}_2 \sigma^y_3 \sigma^x_4 =
-i  \lambda_{1} \hat u_{12} \hat u_{23} \hat u_{34}\lambda_{4} 
\end{equation} (cf.~Fig.~\ref{fig: plaquette_operator}) modify the Majorana band structure. They renormalize the non-universal velocity and change band curvature away from the gapless points, i.e., they contribute irrelevant higher-derivative terms. Despite their simplicity, terms of this type already qualitatively change the properties of vacancies, as we will discuss.

\textit{2. Flux-conserving interactions between $\lambda$ fermions.} Slightly modified multi-spin terms result in four-fermion interactions, i.e., \begin{equation}\delta H_{4} \sim \sigma^y_1 \sigma^{y}_2 \sigma^x_3 \sigma^x_4 =-  \lambda_{1} \hat u_{12} \lambda_{2}\lambda_{3} \hat u_{34}\lambda_{4}~.
\end{equation} In the bulk, they are irrelevant due to the vanishing density of states at the Dirac point.  We will show that such terms play an important role when the number of vacancies on two sublattices is unequal.

\textit{3. Flux-changing terms.} The most generic symmetry-allowed interactions, such as the widely studied Heisenberg and $\Gamma$ terms, fail to commute with $\hat W_\text{\varhexagon}$, i.e., spoil flux conservation. They render the $\mathbb{Z}_2$ gauge field dynamical but do not destabilize the phase until the flux gap closes \cite{Chaloupka2013,Song2016,Gotfryd2017,Gohlke2017,Zhang2021}. These terms have the most dramatic consequences and qualitatively change bulk observables \cite{Song2016}.

Any Kitaev QSL in a real material presumably includes all three of these perturbations. In particular, the first two types of contributions will be generated from the last in models that, microscopically, contain only two-spin couplings. Simulating a generic gapless model on a large two-dimensional system would be a daunting task, which, fortunately, is unneeded for our purposes. We emphasize that all the effects that we describe relate to quantities that are strictly but accidentally zero in the honeycomb model, i.e., they are not required to vanish due to its global symmetries. Consequently, we apply the standard field-theory perspective, asserting that all symmetry-allowed interactions will be generated. If we find a specific interaction that causes an accidentally-zero quantity to become non-zero, adding the most generic terms will not bring them back to zero.

{\bf Vacancies in the honeycomb model.}
To create a vacancy, we now remove the site $\boldsymbol{R}$. The three adjacent plaquette operators  $W_\text{\varhexagon}^{x,y,z}$ are also affected: Removing $\vec \sigma_{\boldsymbol{R}}$  yields three conserved five-spin operators $\tau^\mu_{\boldsymbol{R}}  = \sigma^\mu_{\boldsymbol{R}} W_\text{\varhexagon}^{\mu}$. However, they do not mutually commute. Instead, they satisfy the SU(2) algebra
\begin{align}
[\tau^\mu_{\boldsymbol{R}} ,  \tau^\nu_{\boldsymbol{R}}  ]&= 2i \epsilon^{\mu\nu\kappa}\tau^\kappa_{\boldsymbol{R}}  W_\text{\vacflux}~,
\label{tausu2}
\end{align}
with $W_\text{\vacflux}=\pm 1$ conserved flux around the vacancy site \cite{Dhochak2010}. The $\tau$ operators associated with different vacancies are independent and commute. Consequently, the exact ground state degeneracy is bounded as $g \geq 2^{N}$. 

The actual degeneracy is significantly larger, which is the first result of this work. To obtain it, we note that removing the $\vect R$ site strips three $\xi$ fermions on the adjacent sites of their partner. We denote them by $\psi^\mu_{\vect R}$.
 \begin{figure}
   \centering
\includegraphics[width=.99\linewidth]{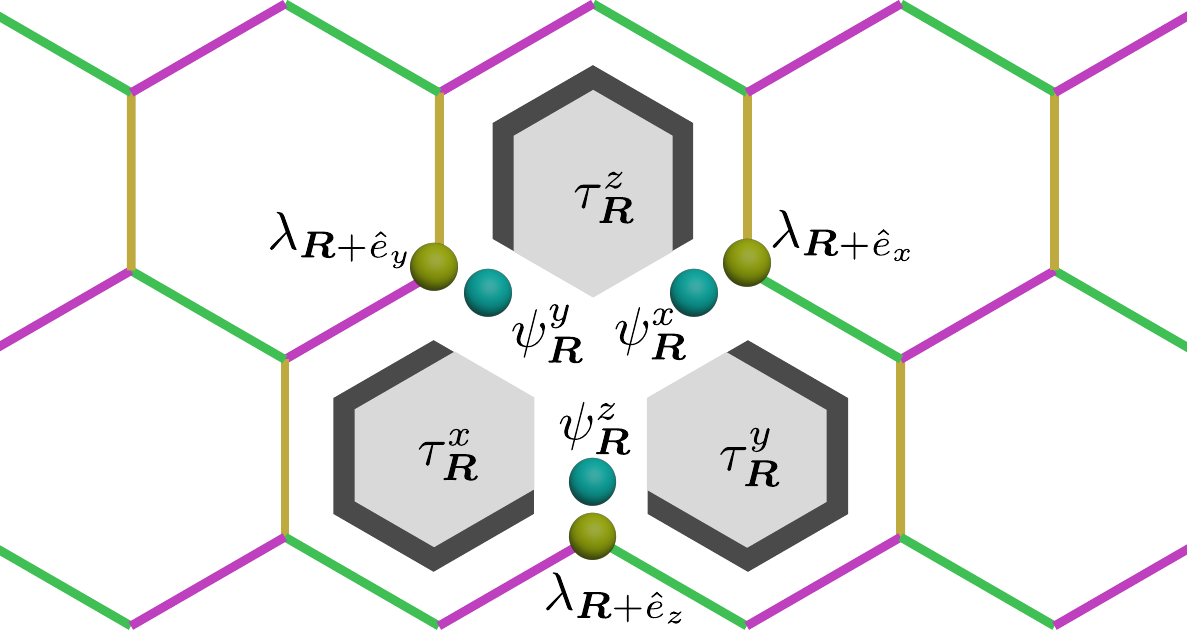}
   \caption{Removing the site $\vect R$ affects three hexagonal plaquettes. The corresponding fluxes are reduced to five-spin operators $\tau^{x,y,z}_{\boldsymbol{R}}$. They are conserved and satisfy the SU(2) algebra Eq.~\eqref{tausu2}. In the fermion representation, one `gauge Majorana' on each of the adjacent sites loses its partner. These dangling Majoranas form a zero-energy subspace and represent the $\tau$ operators according to Eq.~\eqref{vacanymajoranas}.}
   \label{fig: vacancy_notations}
 \end{figure}
Additionally, there is an exact zero-energy mode $\psi^0_{\vect R}$ for the $\lambda$ fermions on the sublattice opposite to that of the vacancy. For any gauge choice of the link variables $\hat u_{\vect  r \vect r'}=\pm 1$, the four zero energy modes $\psi^{0,x,y,z}_{\vect R}$ satisfy
\begin{align}
\vec \tau_{\boldsymbol{R}} =  \pm  \textstyle{\frac{i}{2}}\vec \psi_{\boldsymbol{R}}\times \vec \psi_{\boldsymbol{R}}~, \qquad \psi^0_{\boldsymbol{R}}\psi^x_{\boldsymbol{R}}\psi^y_{\boldsymbol{R}}\psi^z_{\boldsymbol{R}} = \pm 1~.
\label{vacanymajoranas}
\end{align}
The signs in both expressions are not gauge invariant, but they are unimportant for our purposes. For $N_A$ and $N_B$ vacancies on the two sublattices, there are $4(N_A+N_B)$ fermions $\psi^\mu_{\vect R_i}$ subject to a single constraint. However, not all $\psi^0_{\vect R_i}$ represent zero-energy modes; they hybridize pairwise when occupying opposite sublattices. As such, we find that the degeneracy is
\begin{equation}
    g_K = 2^{N_A+N_B + \max{(N_A,N_B)}-1}~,
    \label{eq: degeneracy_formula}
\end{equation}
which is parametrically larger than the bound obtained by counting $\vec \tau$ operators. A source of additional degeneracies are open strings of spin operators that connect two vacancies. Such strings commute with all $W_\text{\varhexagon}$, $\vec \tau$ and are time-reversal odd for vacancies on the same sublattice.

To conclude the summary of vacancies in $H_K$, we consider the vacancy-induced moment. Any $\sigma_{\vect r}$ not adjacent to a vacancy anti-commutes with some of the $W_\text{\varhexagon}$ and thus has zero expectation value. By contrast, $\sigma^\mu_{\boldsymbol{R}+\hat e_\mu}= i \lambda _{\boldsymbol{R}+\hat e_\mu}  \psi^\mu_{\boldsymbol{R}}$ commute with all remaining $W_\text{\varhexagon}$ and could thus acquire a non-zero value in the ground state manifold. Projecting these operators into the degenerate subspace yields 
\begin{align}
P\sigma ^\mu_{\boldsymbol{R}+\hat e_\mu}P &= i {\cal N}^0_{\boldsymbol{R}+\hat e_\mu}\psi^0_{\boldsymbol{R}}  \psi^\mu_{\boldsymbol{R}} = {\cal N}^0_{\boldsymbol{R}+\hat e_\mu}\tau^\mu_{\boldsymbol{R}}~.
\end{align}
Here, ${\cal N}^0_{\boldsymbol{r}}$ is the amplitude of the delocalized zero-mode wave function on the site $\vect r$. 
In the gapless QSL, the zero mode is not normalizable, and $\lim \limits_{L\to \infty}{\cal N}^0(L)= 0$. Consequently, there is no magnetic moment, although the `phantom spin' $\tau$, which has the same global symmetries, has a unit magnitude. 

When the flux around the vacancy is $W_\text{\vacflux}=1$, the zero-mode wave function decays as $1/r$, \cite{Pereira2006,Willans2011}
and the normalization is ${\cal N}_0 \sim \left(\log  L\right)^{-1/2} $. A weak magnetic field $h$ represents a length scale $L_h \sim 1/h$, hence, for $L_h \ll L$, the magnetization becomes $m(h) \sim \left(\log h\right)^{-1/2}$, as found in Ref. \cite{Willans2010}. For $W_\text{\vacflux}=-1$, the zero mode is localized even more poorly.
In either case, the zero-mode amplitude and the vacancy-induced moment are zero in the thermodynamic limit. We have numerically verified that adding moderate bond disorder, with amplitude up to order $|J|$, does not affect the power-law delocalization of zero modes in either flux sector \cite{supplementaryMaterial}. 

{\bf Single vacancy in a generic Kitaev QSL}. When the honeycomb model $H_\text{K}$ of Eq.~\eqref{hkitaev} is perturbed by generic interactions, the behavior of vacancies changes qualitatively. Most significantly, the phantom spins are revealed, and vacancies exhibit a non-zero magnetic moment, which could be observed experimentally. To demonstrate this feature, we consider the local coupling 
\begin{align} H_{ \sigma \tau} = g\sum\nolimits_\mu  \tau^\mu_{\vect R} \sigma^\mu_{\boldsymbol{R}+\hat e_\mu}~,
\label{eqn.toy} 
\end{align}
which preserves the global symmetries of the honeycomb model. It corresponds to a six-spin interaction, which is unlikely to be realized microscopically. Still, it will be generated by generic two-spin terms and must, therefore, be included in the low-energy theory. The analysis of $H=H_\text{K}+H_{\sigma\tau}$ is straightforward: We note that $H$ commutes with all flux operators $\hat W$ and with $\tau^\mu_{\vect R}$. Consequently, any state in the two-fold degenerate ground state manifold corresponds to a specific orientation of the vector $\vec \tau_{\vect R}$ on the Bloch sphere. For any choice of this direction, $H_{\sigma\tau}$ amounts to a Zeeman magnetic field near the vacancy, which induces a non-zero moment, as discussed above.

 \begin{figure}[h]
   \centering
\includegraphics[width=.99\linewidth]{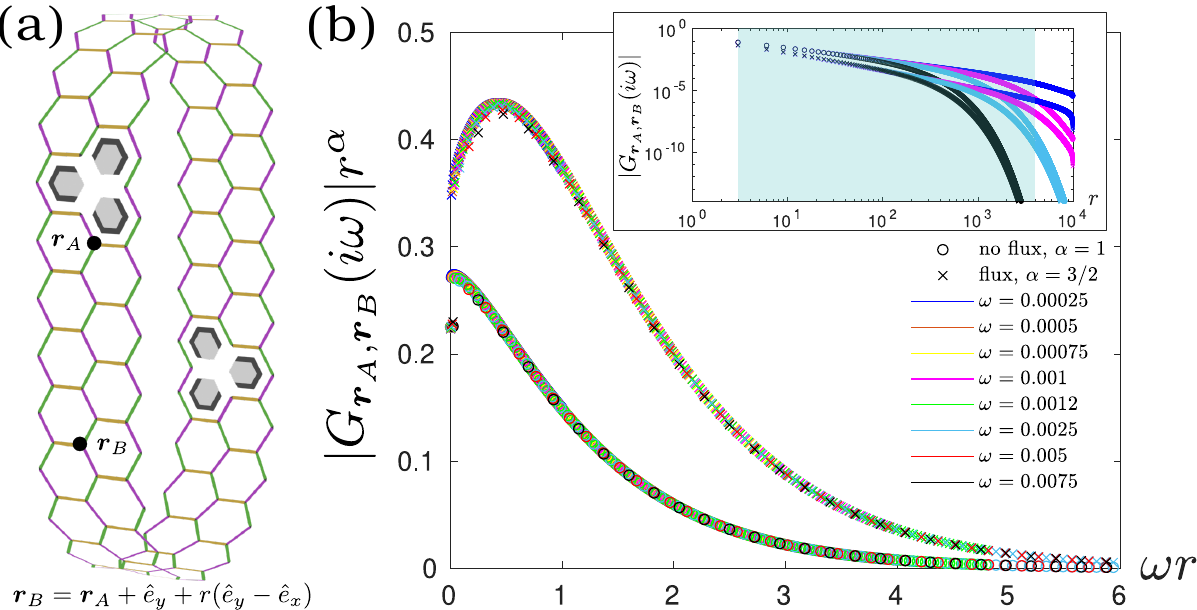}
   \caption{Trapped fluxes modify the scaling behavior of the fermion Green function. We numerically computed the Green function on an infinite cylinder with vacancies on opposite sites. Their separation is $L=10^4$ unit cells. One coordinate is at a fixed position near a vacancy, and the second is varied, as indicated in (a). For $1\ll r,\omega^{-1} \ll L$ the data collapse onto the scaling of Eq.~\eqref{greenscaling} with $z=1$ and $\alpha=1$ or $\alpha=3/2$, depending on the trapped flux (b). A double-logarithmic plot of the bare Green functions is shown in the inset. The shaded region indicates the data range used in the collapse.}
   \label{fig: one_vacancy_data_collapse}
 \end{figure}
The toy model provides a useful proof of principle that generic vacancies are associated with a non-zero magnetic moment. To compute the magnetization profile in a generic situation, we follow the perturbative approach of Ref.~\cite{Song2016}. It demonstrated that the magnetization acquires a contribution from the flux-conserving operator $\delta\sigma_{\vect r}^\mu \sim i f^\mu_{\vect r\vect r'\vect r''}  \lambda_{\vect r'}\lambda_{\vect r''}$, with $\vect r',\vect r''$ on the same sublattice. This contribution is absent in the original honeycomb model $H_\text{K}$ but dominates the long-wavelength behavior in the generic case. Additionally, the `vacancy-Majoranas' $\psi$ of Eq.~\eqref{vacanymajoranas} can couple linearly to the $\lambda$ via 
\begin{align}H_{\psi\lambda} =i t \sum\nolimits_{\mu}\psi_{\vect R}^\mu \lambda_{\vect R_{\mu}}~.
\label{eqn.hybridize}
\end{align}
Here $\vect r_{\mu}$ can be any site on the same sublattice as the vacancy. (It is conceptually simplest, but not necessary, to take $\vect{R}_{x,y,z}$ all different, e.g., by the choice $\vect R_{\mu} = \vect R + 3 \hat e_\mu$). At the second order in $t$ and the first order in $f$, the magnetization at $\vect r$ is given by
\begin{align}
\langle \sigma_{\vect r}\rangle \propto 
\int d\omega G_{\vect r',\vect R_\mu}(\omega)G_{\vect r'',\vect R_{\mu'}}(-\omega)
\langle \tau_{\vect R}\rangle~.
\end{align}
The Green function $G_{\vect r,\vect r'}(\omega) \equiv\langle i \lambda_{\vect r}\lambda_{\vect r'}\rangle_\omega$ at long distances follows the scaling \begin{align}G_{\vect r,\vect r'}(\omega) \sim \omega^\alpha F(\omega|\vect r-\vect r'|^z)~.\label{greenscaling}\end{align} Without vacancies, the exponents are $z=\alpha=1$, and the scaling function $F$ is a modified Bessel function.
When either $\vect r$ or $\vect r'$ is near a vacancy, the scaling is modified depending on the flux at the vacancy. By solving the free-fermion system numerically, we find that $z=1$ remains unchanged. The exponent $\alpha$ remains unity in the absence of flux but changes to  $\alpha^\text{flux}=3/2$ in its presence, see Fig.~\ref{fig: one_vacancy_data_collapse}. Consequently we find $\sigma_{\boldsymbol{r}}\sim |\boldsymbol{r}|^{-3}$ and $\sigma_{\boldsymbol{r}}\sim |\boldsymbol{r}|^{-4}$ for $W_\text{\vacflux}=+1$ and $W_\text{\vacflux}=-1$, respectively. Remarkably, the magnetization around a vacancy can be used to determine the presence of a trapped flux.

\begin{figure}
   \centering \includegraphics[width=0.99\linewidth]{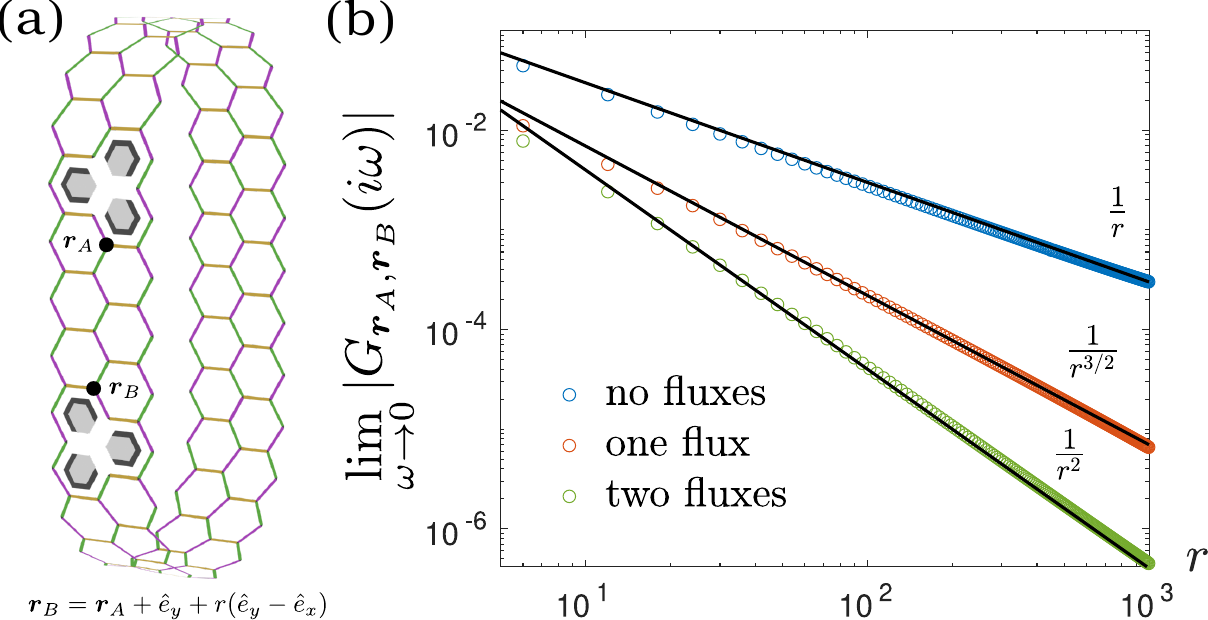}
   \caption{The free-fermion Green functions between the sites adjacent to two vacancies decay with power laws that depend on the trapped fluxes. We computed them numerically for an infinite cylinder with $L=10^4$ unit cells in circumference and for the frequency $\omega=10^{-8} \ll L^{-1}$. The vacancy locations are shown in (a); we increase their separation in steps of three unit cells, the periodicity due to the Dirac points' location (see \refcite{supplementaryMaterial} for details). For all three flux configurations we find distinct power laws, shown in the logarithmic scale in (b). The shown data corresponds to $r=0$ mod $3$ distances between the two sites.}
   \label{fig: two_vacancies_data_scaling}
 \end{figure}

{\bf Multiple vacancies in a generic Kitaev QSL}. We now consider two well-isolated vacancies on opposite sublattices. The delocalized would-be zero-modes hybridize in this case, and the ground state manifold is spanned by the dangling modes alone. They are subject to the constraint \begin{equation}
\psi^{x}_{\vect R}\psi^{y}_{\vect R}\psi^{z}_{\vect R}
\psi^{x}_{\vect R'}\psi^{y}_{\vect R'}\psi^{z}_{\vect R'} = \pm i~,\end{equation}
depending on the choice of the link variables $\hat u$. The constraint implies that $
\tau_{\vect R}^\mu \tau_{\vect R'}^{\nu} =\pm 
 i\psi^\mu_{\boldsymbol{R}}
\psi^{\nu'}_{\boldsymbol{R}'}~$,
i.e., the interaction between phantom spins is a fermion bilinear. When the dangling modes hybridize with the bulk fermions via $H_{\psi \lambda} $ of Eq.~\eqref{eqn.hybridize}, the phantom spins experience an effective coupling \begin{align}
\delta H^{AB}_{\tau\tau} &= t^2 G_{\vect R_{\mu},\vect R'_{\nu}}(\omega=0)\ \tau_{\vect R}^\mu \tau_{\vect R'}^{\nu}~.
\label{htautau}
\end{align}
Notice that this RKKY-like interaction is mediated by a single fermion, unlike its analog in conventional metals, where it arises from particle-hole pairs. Similar to the single-vacancy case, the Green function depends on the fluxes at the vacancies. We numerically computed it for free fermions on an infinite cylinder. Our results show distinct power laws (\figref{fig: two_vacancies_data_scaling}). The Green function decays as $G \sim R^{-1}$, $R^{-3/2}$, or $R^{-2}$ for trapped fluxes at neither, one, or both vacancies, respectively. This scaling remains valid in the presence of additional vacancy pairs (see \refcite{supplementaryMaterial}).

When two vacancies are on the same sublattice, their delocalized zero modes are protected from hybridizing. Indeed, \textit{any} bilinear of $\psi$ fermions is odd under time-reversal symmetry in this case. Still, interactions such as Eq.~\eqref{eqn.toy} with $ PH_{\sigma\tau}P\propto \psi^{x}_{\vect R}\psi^{y}_{\vect R}\psi^{z}_{\vect R}\psi^{0}_{\vect R}$ can reduce the ground state degeneracy to the one encoded by the phantom spins $\vec\tau_{\vect R},\vec\tau_{\vect R'}$. These, in turn, can interact via a more conventional RKKY-like interaction
\begin{align}
\delta H^{AA}_{\tau\tau} &\propto \tau_{\vect R}^\mu \tau_{\vect R'}^{\mu'}\times  \int_\omega G_{\vect R_{\kappa},\vect R'_{\kappa'}}(\omega)
G_{\vect R_{\nu},\vect R'_{\nu'}}(-\omega)~,
\end{align}
when $\psi$ are coupled to the bulk via Eq.~\eqref{eqn.hybridize}.
Based on the scaling determined above, we conclude that the interaction decays as $R^{-3}$, $R^{-4}$ or $R^{-5}$, depending on $W_\text{\vacflux}$ at either vacancy.

To support our perturbative analysis, we used exact diagonalization on a 24-site cluster with two vacancies (see \refcite{supplementaryMaterial}). When the vacancies are on opposite sublattices, we observed that perturbing $H_\text{K}$ with the free-fermion term of Eq.~\eqref{eqn.hybridize} is sufficient to obtain a unique ground state. By contrast, for vacancies on the same sublattice, the same term does not reduce the eight-fold ground-state degeneracy. Further including the four-fermion coupling Eq.~\eqref{eqn.toy} lifts this degeneracy, as we predicted. Additionally, we confirmed that the Kitaev-$\Gamma$-Heisenberg model \cite{PhysRevLett.105.027204,PhysRevLett.112.077204,Chaloupka2013} has a unique ground state in both cases.

{\bf Discussion}. We have shown that generic, symmetry-preserving perturbations to the honeycomb model qualitatively change the properties of vacancies. Notably, isolated vacancies display a non-zero magnetic moment, which could be measured using sensitive magnetometric tools like superconducting quantum interference devices (SQUIDs). Similarly, the bulk magnetic susceptibility $\chi$ for a finite vacancy concentration $n_\text{imp}$ is significantly modified. For magnetic fields $h \gg \sqrt{n_\text{imp}}$, the presence of vacancies in the honeycomb model gives rise to a weakly singular contribution to the magnetic susceptibility characterized by $\chi_\text{imp}(h) \sim n_\text{imp} \log(h) $ \cite{Willans2011}. In contrast, we predict that a generic Kitaev QSL exhibits field-independent magnetization under these conditions. We expect the weak-field limit to display rich physics depending on the vacancy distribution, which would be an interesting topic for future studies.

We note that applying a weak magnetic field to the honeycomb model \eqref{hkitaev} also promotes power-law correlations with amplitude $\propto h^2$ at intermediate length scales $r \ll \Delta^{-1}$, with the gap $\Delta \propto h^{3}$ \cite{Tikhonov2011}. In the generic Kitaev QSL, the dependence changes to $\Delta \propto h$, but the power-law spin correlations become $h$-independent. Here, an isolated vacancy exhibits a moment with singular $h$ dependence \cite{Willans2010}. However, any magnetization induced from this moment via spin correlations with amplitude of order $h^2$ is subleading to the bulk contribution, which is linear.

To connect our results to microscopic spin models, we relate the effective parameters $t,g,f$ to the commonly used Heisenberg-$\Gamma$-$\Gamma'$ terms in the perturbative limit \cite{Hwang2022}.  Specifically, with $\text{Heisenberg-}\Gamma$ terms only, we obtain $g\sim J_{\Gamma}^4/J^3$ or $ J_{\text{Heis.}}^3 J_{\Gamma}^3/J^5,$ and $t,f\sim J_{\text{Heis.}}^2 J_{\Gamma}^2/J^3$, while the inclusion of $\Gamma'$ leads to $t, f \sim J_{\Gamma'}$ \cite{supplementaryMaterial}. We note that, according to \refcite{Zhang2021}, the KSL extends up to $|J_{\text{Heis.}} J_{\Gamma}/J^2|\leq 0.03$ and $|J_{\Gamma}/J| \leq 0.4$.

Vacancies in QSL are closely related to magnetic impurities, i.e., Kondo-type models \cite{Dhochak2010,PhysRevLett.117.037202,PhysRevB.94.024411}.
Our findings regarding how scaling is altered by flux excitations bear direct implications for the RKKY interaction \cite{Dhochak2010}. Furthermore, the interactions among vacancies in a generic QSL may give rise to intriguing multi-impurity Kondo phenomena \cite{PhysRevLett.47.737,PhysRevLett.58.843}. Given that vacancy-vacancy couplings already manifest at the Gaussian level [cf.~Eq.~\eqref{htautau}], it would even be possible to explore them numerically using a free-fermion bath.

Two close cousins of the honeycomb model introduced in Refs.~\cite{Yao2007,Yao2011} also host QSL ground states. They are solvable using a similar fermion representation as in Eq.~\eqref{eq.free}. Moreover, vacancies again host dangling Majorana modes that form locally conserved operators analogous to $\tau^\mu_{\vect R}$, but no magnetization. Adjusting the arguments from the main text, we readily infer that generic perturbations again produce a non-zero moment \cite{supplementaryMaterial}. In the model of Ref. \cite{Yao2007}, time-reversal symmetry is broken spontaneously and determines the orientation of the ensuing moment. In the model of Ref.~\cite{Yao2011}, the extensive ground state degeneracy due to vacancies is lifted via more conventional RKKY interactions.
\\
\begin{acknowledgments}
{\bf Acknowledgments}. It is a pleasure to acknowledge illuminating conversations with Jason Alicea and Elio K\"onig. IY acknowledges the hospitality of the Institute for Theoretical Physics in Cologne and support from the CRC 183 (funded by the Deutsche Forschungsgemeinschaft – project number 277101999) in the final stages of this work. This work was supported by the Israel Science Foundation (ISF) under
grant 2572/21 and by the
Minerva Foundation with funding from the Federal German
Ministry for Education and Research.
\end{acknowledgments}

\bibliographystyle{apsrev4-2}
\bibliography{ref}

\begin{thebibliography}{56}%
\makeatletter
\providecommand \@ifxundefined [1]{%
 \@ifx{#1\undefined}
}%
\providecommand \@ifnum [1]{%
 \ifnum #1\expandafter \@firstoftwo
 \else \expandafter \@secondoftwo
 \fi
}%
\providecommand \@ifx [1]{%
 \ifx #1\expandafter \@firstoftwo
 \else \expandafter \@secondoftwo
 \fi
}%
\providecommand \natexlab [1]{#1}%
\providecommand \enquote  [1]{``#1''}%
\providecommand \bibnamefont  [1]{#1}%
\providecommand \bibfnamefont [1]{#1}%
\providecommand \citenamefont [1]{#1}%
\providecommand \href@noop [0]{\@secondoftwo}%
\providecommand \href [0]{\begingroup \@sanitize@url \@href}%
\providecommand \@href[1]{\@@startlink{#1}\@@href}%
\providecommand \@@href[1]{\endgroup#1\@@endlink}%
\providecommand \@sanitize@url [0]{\catcode `\\12\catcode `\$12\catcode `\&12\catcode `\#12\catcode `\^12\catcode `\_12\catcode `\%12\relax}%
\providecommand \@@startlink[1]{}%
\providecommand \@@endlink[0]{}%
\providecommand \url  [0]{\begingroup\@sanitize@url \@url }%
\providecommand \@url [1]{\endgroup\@href {#1}{\urlprefix }}%
\providecommand \urlprefix  [0]{URL }%
\providecommand \Eprint [0]{\href }%
\providecommand \doibase [0]{https://doi.org/}%
\providecommand \selectlanguage [0]{\@gobble}%
\providecommand \bibinfo  [0]{\@secondoftwo}%
\providecommand \bibfield  [0]{\@secondoftwo}%
\providecommand \translation [1]{[#1]}%
\providecommand \BibitemOpen [0]{}%
\providecommand \bibitemStop [0]{}%
\providecommand \bibitemNoStop [0]{.\EOS\space}%
\providecommand \EOS [0]{\spacefactor3000\relax}%
\providecommand \BibitemShut  [1]{\csname bibitem#1\endcsname}%
\let\auto@bib@innerbib\@empty
\bibitem [{\citenamefont {Balents}(2010)}]{Balents2010}%
  \BibitemOpen
  \bibfield  {author} {\bibinfo {author} {\bibfnamefont {L.}~\bibnamefont {Balents}},\ }\href {https://doi.org/10.1038/nature08917} {\bibfield  {journal} {\bibinfo  {journal} {Nature}\ }\textbf {\bibinfo {volume} {464}},\ \bibinfo {pages} {199} (\bibinfo {year} {2010})}\BibitemShut {NoStop}%
\bibitem [{\citenamefont {Savary}\ and\ \citenamefont {Balents}(2016)}]{Savary2016}%
  \BibitemOpen
  \bibfield  {author} {\bibinfo {author} {\bibfnamefont {L.}~\bibnamefont {Savary}}\ and\ \bibinfo {author} {\bibfnamefont {L.}~\bibnamefont {Balents}},\ }\href {https://doi.org/10.1088/0034-4885/80/1/016502} {\bibfield  {journal} {\bibinfo  {journal} {Reports on Progress in Physics}\ }\textbf {\bibinfo {volume} {80}},\ \bibinfo {pages} {016502} (\bibinfo {year} {2016})}\BibitemShut {NoStop}%
\bibitem [{\citenamefont {Norman}(2016)}]{Norman2016}%
  \BibitemOpen
  \bibfield  {author} {\bibinfo {author} {\bibfnamefont {M.~R.}\ \bibnamefont {Norman}},\ }\href {https://doi.org/10.1103/RevModPhys.88.041002} {\bibfield  {journal} {\bibinfo  {journal} {Rev. Mod. Phys.}\ }\textbf {\bibinfo {volume} {88}},\ \bibinfo {pages} {041002} (\bibinfo {year} {2016})}\BibitemShut {NoStop}%
\bibitem [{\citenamefont {Zhou}\ \emph {et~al.}(2017)\citenamefont {Zhou}, \citenamefont {Kanoda},\ and\ \citenamefont {Ng}}]{Zhou2017}%
  \BibitemOpen
  \bibfield  {author} {\bibinfo {author} {\bibfnamefont {Y.}~\bibnamefont {Zhou}}, \bibinfo {author} {\bibfnamefont {K.}~\bibnamefont {Kanoda}},\ and\ \bibinfo {author} {\bibfnamefont {T.-K.}\ \bibnamefont {Ng}},\ }\href {https://doi.org/10.1103/RevModPhys.89.025003} {\bibfield  {journal} {\bibinfo  {journal} {Rev. Mod. Phys.}\ }\textbf {\bibinfo {volume} {89}},\ \bibinfo {pages} {025003} (\bibinfo {year} {2017})}\BibitemShut {NoStop}%
\bibitem [{\citenamefont {{Takagi}}\ \emph {et~al.}(2019)\citenamefont {{Takagi}}, \citenamefont {{Takayama}}, \citenamefont {{Jackeli}}, \citenamefont {{Khaliullin}},\ and\ \citenamefont {{Nagler}}}]{Takagi2019}%
  \BibitemOpen
  \bibfield  {author} {\bibinfo {author} {\bibfnamefont {H.}~\bibnamefont {{Takagi}}}, \bibinfo {author} {\bibfnamefont {T.}~\bibnamefont {{Takayama}}}, \bibinfo {author} {\bibfnamefont {G.}~\bibnamefont {{Jackeli}}}, \bibinfo {author} {\bibfnamefont {G.}~\bibnamefont {{Khaliullin}}},\ and\ \bibinfo {author} {\bibfnamefont {S.~E.}\ \bibnamefont {{Nagler}}},\ }\href {https://doi.org/10.1038/s42254-019-0038-2} {\bibfield  {journal} {\bibinfo  {journal} {Nature Reviews Physics}\ }\textbf {\bibinfo {volume} {1}},\ \bibinfo {pages} {264} (\bibinfo {year} {2019})}\BibitemShut {NoStop}%
\bibitem [{\citenamefont {Knolle}\ and\ \citenamefont {Moessner}(2019)}]{Knolle2019}%
  \BibitemOpen
  \bibfield  {author} {\bibinfo {author} {\bibfnamefont {J.}~\bibnamefont {Knolle}}\ and\ \bibinfo {author} {\bibfnamefont {R.}~\bibnamefont {Moessner}},\ }\href {https://doi.org/10.1146/annurev-conmatphys-031218-013401} {\bibfield  {journal} {\bibinfo  {journal} {Annu. Rev. Condens. Matter Phys.}\ }\textbf {\bibinfo {volume} {10}},\ \bibinfo {pages} {451} (\bibinfo {year} {2019})}\BibitemShut {NoStop}%
\bibitem [{\citenamefont {Broholm}\ \emph {et~al.}(2020)\citenamefont {Broholm}, \citenamefont {Cava}, \citenamefont {Kivelson}, \citenamefont {Nocera}, \citenamefont {Norman},\ and\ \citenamefont {Senthil}}]{Broholm2020}%
  \BibitemOpen
  \bibfield  {author} {\bibinfo {author} {\bibfnamefont {C.}~\bibnamefont {Broholm}}, \bibinfo {author} {\bibfnamefont {R.~J.}\ \bibnamefont {Cava}}, \bibinfo {author} {\bibfnamefont {S.~A.}\ \bibnamefont {Kivelson}}, \bibinfo {author} {\bibfnamefont {D.~G.}\ \bibnamefont {Nocera}}, \bibinfo {author} {\bibfnamefont {M.~R.}\ \bibnamefont {Norman}},\ and\ \bibinfo {author} {\bibfnamefont {T.}~\bibnamefont {Senthil}},\ }\href {https://doi.org/10.1126/science.aay0668} {\bibfield  {journal} {\bibinfo  {journal} {Science}\ }\textbf {\bibinfo {volume} {367}},\ \bibinfo {pages} {eaay0668} (\bibinfo {year} {2020})}\BibitemShut {NoStop}%
\bibitem [{\citenamefont {Kitaev}(2006)}]{Kitaev2006}%
  \BibitemOpen
  \bibfield  {author} {\bibinfo {author} {\bibfnamefont {A.}~\bibnamefont {Kitaev}},\ }\href {https://doi.org/https://doi.org/10.1016/j.aop.2005.10.005} {\bibfield  {journal} {\bibinfo  {journal} {Annals of Physics}\ }\textbf {\bibinfo {volume} {321}},\ \bibinfo {pages} {2} (\bibinfo {year} {2006})},\ \bibinfo {note} {january Special Issue}\BibitemShut {NoStop}%
\bibitem [{\citenamefont {{Banerjee}}\ \emph {et~al.}(2016)\citenamefont {{Banerjee}}, \citenamefont {{Bridges}}, \citenamefont {{Yan}}, \citenamefont {{Aczel}}, \citenamefont {{Li}}, \citenamefont {{Stone}}, \citenamefont {{Granroth}}, \citenamefont {{Lumsden}}, \citenamefont {{Yiu}}, \citenamefont {{Knolle}}, \citenamefont {{Bhattacharjee}}, \citenamefont {{Kovrizhin}}, \citenamefont {{Moessner}}, \citenamefont {{Tennant}}, \citenamefont {{Mandrus}},\ and\ \citenamefont {{Nagler}}}]{Banerjee2016}%
  \BibitemOpen
  \bibfield  {author} {\bibinfo {author} {\bibfnamefont {A.}~\bibnamefont {{Banerjee}}}, \bibinfo {author} {\bibfnamefont {C.~A.}\ \bibnamefont {{Bridges}}}, \bibinfo {author} {\bibfnamefont {J.~Q.}\ \bibnamefont {{Yan}}}, \bibinfo {author} {\bibfnamefont {A.~A.}\ \bibnamefont {{Aczel}}}, \bibinfo {author} {\bibfnamefont {L.}~\bibnamefont {{Li}}}, \bibinfo {author} {\bibfnamefont {M.~B.}\ \bibnamefont {{Stone}}}, \bibinfo {author} {\bibfnamefont {G.~E.}\ \bibnamefont {{Granroth}}}, \bibinfo {author} {\bibfnamefont {M.~D.}\ \bibnamefont {{Lumsden}}}, \bibinfo {author} {\bibfnamefont {Y.}~\bibnamefont {{Yiu}}}, \bibinfo {author} {\bibfnamefont {J.}~\bibnamefont {{Knolle}}}, \bibinfo {author} {\bibfnamefont {S.}~\bibnamefont {{Bhattacharjee}}}, \bibinfo {author} {\bibfnamefont {D.~L.}\ \bibnamefont {{Kovrizhin}}}, \bibinfo {author} {\bibfnamefont {R.}~\bibnamefont {{Moessner}}}, \bibinfo {author} {\bibfnamefont {D.~A.}\ \bibnamefont {{Tennant}}}, \bibinfo {author} {\bibfnamefont {D.~G.}\ \bibnamefont
  {{Mandrus}}},\ and\ \bibinfo {author} {\bibfnamefont {S.~E.}\ \bibnamefont {{Nagler}}},\ }\href {https://doi.org/10.1038/nmat4604} {\bibfield  {journal} {\bibinfo  {journal} {Nature Materials}\ }\textbf {\bibinfo {volume} {15}},\ \bibinfo {pages} {733} (\bibinfo {year} {2016})}\BibitemShut {NoStop}%
\bibitem [{\citenamefont {Winter}\ \emph {et~al.}(2017)\citenamefont {Winter}, \citenamefont {Tsirlin}, \citenamefont {Daghofer}, \citenamefont {van~den Brink}, \citenamefont {Singh}, \citenamefont {Gegenwart},\ and\ \citenamefont {Valentí}}]{Winter2017}%
  \BibitemOpen
  \bibfield  {author} {\bibinfo {author} {\bibfnamefont {S.~M.}\ \bibnamefont {Winter}}, \bibinfo {author} {\bibfnamefont {A.~A.}\ \bibnamefont {Tsirlin}}, \bibinfo {author} {\bibfnamefont {M.}~\bibnamefont {Daghofer}}, \bibinfo {author} {\bibfnamefont {J.}~\bibnamefont {van~den Brink}}, \bibinfo {author} {\bibfnamefont {Y.}~\bibnamefont {Singh}}, \bibinfo {author} {\bibfnamefont {P.}~\bibnamefont {Gegenwart}},\ and\ \bibinfo {author} {\bibfnamefont {R.}~\bibnamefont {Valentí}},\ }\href {https://doi.org/10.1088/1361-648X/aa8cf5} {\bibfield  {journal} {\bibinfo  {journal} {Journal of Physics: Condensed Matter}\ }\textbf {\bibinfo {volume} {29}},\ \bibinfo {pages} {493002} (\bibinfo {year} {2017})}\BibitemShut {NoStop}%
\bibitem [{\citenamefont {Hermanns}\ \emph {et~al.}(2018)\citenamefont {Hermanns}, \citenamefont {Kimchi},\ and\ \citenamefont {Knolle}}]{Hermanns2018}%
  \BibitemOpen
  \bibfield  {author} {\bibinfo {author} {\bibfnamefont {M.}~\bibnamefont {Hermanns}}, \bibinfo {author} {\bibfnamefont {I.}~\bibnamefont {Kimchi}},\ and\ \bibinfo {author} {\bibfnamefont {J.}~\bibnamefont {Knolle}},\ }\href {https://doi.org/10.1146/annurev-conmatphys-033117-053934} {\bibfield  {journal} {\bibinfo  {journal} {Annual Review of Condensed Matter Physics}\ }\textbf {\bibinfo {volume} {9}},\ \bibinfo {pages} {17} (\bibinfo {year} {2018})}\BibitemShut {NoStop}%
\bibitem [{\citenamefont {Motome}\ and\ \citenamefont {Nasu}(2020)}]{Motome2020}%
  \BibitemOpen
  \bibfield  {author} {\bibinfo {author} {\bibfnamefont {Y.}~\bibnamefont {Motome}}\ and\ \bibinfo {author} {\bibfnamefont {J.}~\bibnamefont {Nasu}},\ }\href {https://doi.org/10.7566/JPSJ.89.012002} {\bibfield  {journal} {\bibinfo  {journal} {Journal of the Physical Society of Japan}\ }\textbf {\bibinfo {volume} {89}},\ \bibinfo {pages} {012002} (\bibinfo {year} {2020})}\BibitemShut {NoStop}%
\bibitem [{\citenamefont {Trebst}\ and\ \citenamefont {Hickey}(2022)}]{Trebst2022}%
  \BibitemOpen
  \bibfield  {author} {\bibinfo {author} {\bibfnamefont {S.}~\bibnamefont {Trebst}}\ and\ \bibinfo {author} {\bibfnamefont {C.}~\bibnamefont {Hickey}},\ }\href {https://doi.org/https://doi.org/10.1016/j.physrep.2021.11.003} {\bibfield  {journal} {\bibinfo  {journal} {Physics Reports}\ }\textbf {\bibinfo {volume} {950}},\ \bibinfo {pages} {1} (\bibinfo {year} {2022})}\BibitemShut {NoStop}%
\bibitem [{\citenamefont {{Kasahara}}\ \emph {et~al.}(2018)\citenamefont {{Kasahara}}, \citenamefont {{Ohnishi}}, \citenamefont {{Mizukami}}, \citenamefont {{Tanaka}}, \citenamefont {{Ma}}, \citenamefont {{Sugii}}, \citenamefont {{Kurita}}, \citenamefont {{Tanaka}}, \citenamefont {{Nasu}}, \citenamefont {{Motome}}, \citenamefont {{Shibauchi}},\ and\ \citenamefont {{Matsuda}}}]{Kasahara2018}%
  \BibitemOpen
  \bibfield  {author} {\bibinfo {author} {\bibfnamefont {Y.}~\bibnamefont {{Kasahara}}}, \bibinfo {author} {\bibfnamefont {T.}~\bibnamefont {{Ohnishi}}}, \bibinfo {author} {\bibfnamefont {Y.}~\bibnamefont {{Mizukami}}}, \bibinfo {author} {\bibfnamefont {O.}~\bibnamefont {{Tanaka}}}, \bibinfo {author} {\bibfnamefont {S.}~\bibnamefont {{Ma}}}, \bibinfo {author} {\bibfnamefont {K.}~\bibnamefont {{Sugii}}}, \bibinfo {author} {\bibfnamefont {N.}~\bibnamefont {{Kurita}}}, \bibinfo {author} {\bibfnamefont {H.}~\bibnamefont {{Tanaka}}}, \bibinfo {author} {\bibfnamefont {J.}~\bibnamefont {{Nasu}}}, \bibinfo {author} {\bibfnamefont {Y.}~\bibnamefont {{Motome}}}, \bibinfo {author} {\bibfnamefont {T.}~\bibnamefont {{Shibauchi}}},\ and\ \bibinfo {author} {\bibfnamefont {Y.}~\bibnamefont {{Matsuda}}},\ }\href {https://doi.org/10.1038/s41586-018-0274-0} {\bibfield  {journal} {\bibinfo  {journal} {\nat}\ }\textbf {\bibinfo {volume} {559}},\ \bibinfo {pages} {227} (\bibinfo {year} {2018})}\BibitemShut {NoStop}%
\bibitem [{\citenamefont {{Kitagawa}}\ \emph {et~al.}(2018)\citenamefont {{Kitagawa}}, \citenamefont {{Takayama}}, \citenamefont {{Matsumoto}}, \citenamefont {{Kato}}, \citenamefont {{Takano}}, \citenamefont {{Kishimoto}}, \citenamefont {{Bette}}, \citenamefont {{Dinnebier}}, \citenamefont {{Jackeli}},\ and\ \citenamefont {{Takagi}}}]{Kitagava2018}%
  \BibitemOpen
  \bibfield  {author} {\bibinfo {author} {\bibfnamefont {K.}~\bibnamefont {{Kitagawa}}}, \bibinfo {author} {\bibfnamefont {T.}~\bibnamefont {{Takayama}}}, \bibinfo {author} {\bibfnamefont {Y.}~\bibnamefont {{Matsumoto}}}, \bibinfo {author} {\bibfnamefont {A.}~\bibnamefont {{Kato}}}, \bibinfo {author} {\bibfnamefont {R.}~\bibnamefont {{Takano}}}, \bibinfo {author} {\bibfnamefont {Y.}~\bibnamefont {{Kishimoto}}}, \bibinfo {author} {\bibfnamefont {S.}~\bibnamefont {{Bette}}}, \bibinfo {author} {\bibfnamefont {R.}~\bibnamefont {{Dinnebier}}}, \bibinfo {author} {\bibfnamefont {G.}~\bibnamefont {{Jackeli}}},\ and\ \bibinfo {author} {\bibfnamefont {H.}~\bibnamefont {{Takagi}}},\ }\href {https://doi.org/10.1038/nature25482} {\bibfield  {journal} {\bibinfo  {journal} {\nat}\ }\textbf {\bibinfo {volume} {554}},\ \bibinfo {pages} {341} (\bibinfo {year} {2018})}\BibitemShut {NoStop}%
\bibitem [{\citenamefont {Takahashi}\ \emph {et~al.}(2019)\citenamefont {Takahashi}, \citenamefont {Wang}, \citenamefont {Arsenault}, \citenamefont {Imai}, \citenamefont {Abramchuk}, \citenamefont {Tafti},\ and\ \citenamefont {Singer}}]{Takahashi2019}%
  \BibitemOpen
  \bibfield  {author} {\bibinfo {author} {\bibfnamefont {S.~K.}\ \bibnamefont {Takahashi}}, \bibinfo {author} {\bibfnamefont {J.}~\bibnamefont {Wang}}, \bibinfo {author} {\bibfnamefont {A.}~\bibnamefont {Arsenault}}, \bibinfo {author} {\bibfnamefont {T.}~\bibnamefont {Imai}}, \bibinfo {author} {\bibfnamefont {M.}~\bibnamefont {Abramchuk}}, \bibinfo {author} {\bibfnamefont {F.}~\bibnamefont {Tafti}},\ and\ \bibinfo {author} {\bibfnamefont {P.~M.}\ \bibnamefont {Singer}},\ }\href {https://doi.org/10.1103/PhysRevX.9.031047} {\bibfield  {journal} {\bibinfo  {journal} {Phys. Rev. X}\ }\textbf {\bibinfo {volume} {9}},\ \bibinfo {pages} {031047} (\bibinfo {year} {2019})}\BibitemShut {NoStop}%
\bibitem [{\citenamefont {Yamada}(2020)}]{Yamada2020}%
  \BibitemOpen
  \bibfield  {author} {\bibinfo {author} {\bibfnamefont {M.~G.}\ \bibnamefont {Yamada}},\ }\href {https://doi.org/10.1038/s41535-020-00285-3} {\bibfield  {journal} {\bibinfo  {journal} {npj Quantum Materials}\ }\textbf {\bibinfo {volume} {5}},\ \bibinfo {pages} {82} (\bibinfo {year} {2020})}\BibitemShut {NoStop}%
\bibitem [{\citenamefont {Do}\ \emph {et~al.}(2020)\citenamefont {Do}, \citenamefont {Lee}, \citenamefont {Kihara}, \citenamefont {Choi}, \citenamefont {Yoon}, \citenamefont {Kim}, \citenamefont {Cheong}, \citenamefont {Chen}, \citenamefont {Chou}, \citenamefont {Nojiri},\ and\ \citenamefont {Choi}}]{Do2020}%
  \BibitemOpen
  \bibfield  {author} {\bibinfo {author} {\bibfnamefont {S.-H.}\ \bibnamefont {Do}}, \bibinfo {author} {\bibfnamefont {C.~H.}\ \bibnamefont {Lee}}, \bibinfo {author} {\bibfnamefont {T.}~\bibnamefont {Kihara}}, \bibinfo {author} {\bibfnamefont {Y.~S.}\ \bibnamefont {Choi}}, \bibinfo {author} {\bibfnamefont {S.}~\bibnamefont {Yoon}}, \bibinfo {author} {\bibfnamefont {K.}~\bibnamefont {Kim}}, \bibinfo {author} {\bibfnamefont {H.}~\bibnamefont {Cheong}}, \bibinfo {author} {\bibfnamefont {W.-T.}\ \bibnamefont {Chen}}, \bibinfo {author} {\bibfnamefont {F.}~\bibnamefont {Chou}}, \bibinfo {author} {\bibfnamefont {H.}~\bibnamefont {Nojiri}},\ and\ \bibinfo {author} {\bibfnamefont {K.-Y.}\ \bibnamefont {Choi}},\ }\href {https://doi.org/10.1103/PhysRevLett.124.047204} {\bibfield  {journal} {\bibinfo  {journal} {Phys. Rev. Lett.}\ }\textbf {\bibinfo {volume} {124}},\ \bibinfo {pages} {047204} (\bibinfo {year} {2020})}\BibitemShut {NoStop}%
\bibitem [{\citenamefont {Yokoi}\ \emph {et~al.}(2021)\citenamefont {Yokoi}, \citenamefont {Ma}, \citenamefont {Kasahara}, \citenamefont {Kasahara}, \citenamefont {Shibauchi}, \citenamefont {Kurita}, \citenamefont {Tanaka}, \citenamefont {Nasu}, \citenamefont {Motome}, \citenamefont {Hickey}, \citenamefont {Trebst},\ and\ \citenamefont {Matsuda}}]{Yokoi2021}%
  \BibitemOpen
  \bibfield  {author} {\bibinfo {author} {\bibfnamefont {T.}~\bibnamefont {Yokoi}}, \bibinfo {author} {\bibfnamefont {S.}~\bibnamefont {Ma}}, \bibinfo {author} {\bibfnamefont {Y.}~\bibnamefont {Kasahara}}, \bibinfo {author} {\bibfnamefont {S.}~\bibnamefont {Kasahara}}, \bibinfo {author} {\bibfnamefont {T.}~\bibnamefont {Shibauchi}}, \bibinfo {author} {\bibfnamefont {N.}~\bibnamefont {Kurita}}, \bibinfo {author} {\bibfnamefont {H.}~\bibnamefont {Tanaka}}, \bibinfo {author} {\bibfnamefont {J.}~\bibnamefont {Nasu}}, \bibinfo {author} {\bibfnamefont {Y.}~\bibnamefont {Motome}}, \bibinfo {author} {\bibfnamefont {C.}~\bibnamefont {Hickey}}, \bibinfo {author} {\bibfnamefont {S.}~\bibnamefont {Trebst}},\ and\ \bibinfo {author} {\bibfnamefont {Y.}~\bibnamefont {Matsuda}},\ }\href {https://doi.org/10.1126/science.aay5551} {\bibfield  {journal} {\bibinfo  {journal} {Science}\ }\textbf {\bibinfo {volume} {373}},\ \bibinfo {pages} {568} (\bibinfo {year} {2021})}\BibitemShut {NoStop}%
\bibitem [{\citenamefont {Bruin}\ \emph {et~al.}(2022)\citenamefont {Bruin}, \citenamefont {Claus}, \citenamefont {Matsumoto}, \citenamefont {Kurita}, \citenamefont {Tanaka},\ and\ \citenamefont {Takagi}}]{Bruin2022}%
  \BibitemOpen
  \bibfield  {author} {\bibinfo {author} {\bibfnamefont {J.~A.~N.}\ \bibnamefont {Bruin}}, \bibinfo {author} {\bibfnamefont {R.~R.}\ \bibnamefont {Claus}}, \bibinfo {author} {\bibfnamefont {Y.}~\bibnamefont {Matsumoto}}, \bibinfo {author} {\bibfnamefont {N.}~\bibnamefont {Kurita}}, \bibinfo {author} {\bibfnamefont {H.}~\bibnamefont {Tanaka}},\ and\ \bibinfo {author} {\bibfnamefont {H.}~\bibnamefont {Takagi}},\ }\href {https://doi.org/10.1038/s41567-021-01501-y} {\bibfield  {journal} {\bibinfo  {journal} {Nature Physics}\ }\textbf {\bibinfo {volume} {18}},\ \bibinfo {pages} {401} (\bibinfo {year} {2022})}\BibitemShut {NoStop}%
\bibitem [{\citenamefont {Czajka}\ \emph {et~al.}(2021)\citenamefont {Czajka}, \citenamefont {Gao}, \citenamefont {Hirschberger}, \citenamefont {Lampen-Kelley}, \citenamefont {Banerjee}, \citenamefont {Yan}, \citenamefont {Mandrus}, \citenamefont {Nagler},\ and\ \citenamefont {Ong}}]{Czajka2021}%
  \BibitemOpen
  \bibfield  {author} {\bibinfo {author} {\bibfnamefont {P.}~\bibnamefont {Czajka}}, \bibinfo {author} {\bibfnamefont {T.}~\bibnamefont {Gao}}, \bibinfo {author} {\bibfnamefont {M.}~\bibnamefont {Hirschberger}}, \bibinfo {author} {\bibfnamefont {P.}~\bibnamefont {Lampen-Kelley}}, \bibinfo {author} {\bibfnamefont {A.}~\bibnamefont {Banerjee}}, \bibinfo {author} {\bibfnamefont {J.}~\bibnamefont {Yan}}, \bibinfo {author} {\bibfnamefont {D.~G.}\ \bibnamefont {Mandrus}}, \bibinfo {author} {\bibfnamefont {S.~E.}\ \bibnamefont {Nagler}},\ and\ \bibinfo {author} {\bibfnamefont {N.~P.}\ \bibnamefont {Ong}},\ }\href {https://doi.org/10.1038/s41567-021-01243-x} {\bibfield  {journal} {\bibinfo  {journal} {Nature Physics}\ }\textbf {\bibinfo {volume} {17}},\ \bibinfo {pages} {915} (\bibinfo {year} {2021})}\BibitemShut {NoStop}%
\bibitem [{\citenamefont {Czajka}\ \emph {et~al.}(2023)\citenamefont {Czajka}, \citenamefont {Gao}, \citenamefont {Hirschberger}, \citenamefont {Lampen-Kelley}, \citenamefont {Banerjee}, \citenamefont {Quirk}, \citenamefont {Mandrus}, \citenamefont {Nagler},\ and\ \citenamefont {Ong}}]{Czajka2023}%
  \BibitemOpen
  \bibfield  {author} {\bibinfo {author} {\bibfnamefont {P.}~\bibnamefont {Czajka}}, \bibinfo {author} {\bibfnamefont {T.}~\bibnamefont {Gao}}, \bibinfo {author} {\bibfnamefont {M.}~\bibnamefont {Hirschberger}}, \bibinfo {author} {\bibfnamefont {P.}~\bibnamefont {Lampen-Kelley}}, \bibinfo {author} {\bibfnamefont {A.}~\bibnamefont {Banerjee}}, \bibinfo {author} {\bibfnamefont {N.}~\bibnamefont {Quirk}}, \bibinfo {author} {\bibfnamefont {D.~G.}\ \bibnamefont {Mandrus}}, \bibinfo {author} {\bibfnamefont {S.~E.}\ \bibnamefont {Nagler}},\ and\ \bibinfo {author} {\bibfnamefont {N.~P.}\ \bibnamefont {Ong}},\ }\href {https://doi.org/10.1038/s41563-022-01397-w} {\bibfield  {journal} {\bibinfo  {journal} {Nature Materials}\ }\textbf {\bibinfo {volume} {22}},\ \bibinfo {pages} {36} (\bibinfo {year} {2023})}\BibitemShut {NoStop}%
\bibitem [{\citenamefont {Chaloupka}\ \emph {et~al.}(2010)\citenamefont {Chaloupka}, \citenamefont {Jackeli},\ and\ \citenamefont {Khaliullin}}]{PhysRevLett.105.027204}%
  \BibitemOpen
  \bibfield  {author} {\bibinfo {author} {\bibfnamefont {J.~c.~v.}\ \bibnamefont {Chaloupka}}, \bibinfo {author} {\bibfnamefont {G.}~\bibnamefont {Jackeli}},\ and\ \bibinfo {author} {\bibfnamefont {G.}~\bibnamefont {Khaliullin}},\ }\href {https://doi.org/10.1103/PhysRevLett.105.027204} {\bibfield  {journal} {\bibinfo  {journal} {Phys. Rev. Lett.}\ }\textbf {\bibinfo {volume} {105}},\ \bibinfo {pages} {027204} (\bibinfo {year} {2010})}\BibitemShut {NoStop}%
\bibitem [{\citenamefont {Rau}\ \emph {et~al.}(2014{\natexlab{a}})\citenamefont {Rau}, \citenamefont {Lee},\ and\ \citenamefont {Kee}}]{PhysRevLett.112.077204}%
  \BibitemOpen
  \bibfield  {author} {\bibinfo {author} {\bibfnamefont {J.~G.}\ \bibnamefont {Rau}}, \bibinfo {author} {\bibfnamefont {E.~K.-H.}\ \bibnamefont {Lee}},\ and\ \bibinfo {author} {\bibfnamefont {H.-Y.}\ \bibnamefont {Kee}},\ }\href {https://doi.org/10.1103/PhysRevLett.112.077204} {\bibfield  {journal} {\bibinfo  {journal} {Phys. Rev. Lett.}\ }\textbf {\bibinfo {volume} {112}},\ \bibinfo {pages} {077204} (\bibinfo {year} {2014}{\natexlab{a}})}\BibitemShut {NoStop}%
\bibitem [{\citenamefont {Jiang}\ \emph {et~al.}(2011)\citenamefont {Jiang}, \citenamefont {Gu}, \citenamefont {Qi},\ and\ \citenamefont {Trebst}}]{Jiang2011}%
  \BibitemOpen
  \bibfield  {author} {\bibinfo {author} {\bibfnamefont {H.-C.}\ \bibnamefont {Jiang}}, \bibinfo {author} {\bibfnamefont {Z.-C.}\ \bibnamefont {Gu}}, \bibinfo {author} {\bibfnamefont {X.-L.}\ \bibnamefont {Qi}},\ and\ \bibinfo {author} {\bibfnamefont {S.}~\bibnamefont {Trebst}},\ }\href {https://doi.org/10.1103/PhysRevB.83.245104} {\bibfield  {journal} {\bibinfo  {journal} {Phys. Rev. B}\ }\textbf {\bibinfo {volume} {83}},\ \bibinfo {pages} {245104} (\bibinfo {year} {2011})}\BibitemShut {NoStop}%
\bibitem [{\citenamefont {Tikhonov}\ \emph {et~al.}(2011)\citenamefont {Tikhonov}, \citenamefont {Feigel'man},\ and\ \citenamefont {Kitaev}}]{Tikhonov2011}%
  \BibitemOpen
  \bibfield  {author} {\bibinfo {author} {\bibfnamefont {K.~S.}\ \bibnamefont {Tikhonov}}, \bibinfo {author} {\bibfnamefont {M.~V.}\ \bibnamefont {Feigel'man}},\ and\ \bibinfo {author} {\bibfnamefont {A.~Y.}\ \bibnamefont {Kitaev}},\ }\href {https://doi.org/10.1103/PhysRevLett.106.067203} {\bibfield  {journal} {\bibinfo  {journal} {Phys. Rev. Lett.}\ }\textbf {\bibinfo {volume} {106}},\ \bibinfo {pages} {067203} (\bibinfo {year} {2011})}\BibitemShut {NoStop}%
\bibitem [{\citenamefont {Chaloupka}\ \emph {et~al.}(2013)\citenamefont {Chaloupka}, \citenamefont {Jackeli},\ and\ \citenamefont {Khaliullin}}]{Chaloupka2013}%
  \BibitemOpen
  \bibfield  {author} {\bibinfo {author} {\bibfnamefont {J.~c.~v.}\ \bibnamefont {Chaloupka}}, \bibinfo {author} {\bibfnamefont {G.}~\bibnamefont {Jackeli}},\ and\ \bibinfo {author} {\bibfnamefont {G.}~\bibnamefont {Khaliullin}},\ }\href {https://doi.org/10.1103/PhysRevLett.110.097204} {\bibfield  {journal} {\bibinfo  {journal} {Phys. Rev. Lett.}\ }\textbf {\bibinfo {volume} {110}},\ \bibinfo {pages} {097204} (\bibinfo {year} {2013})}\BibitemShut {NoStop}%
\bibitem [{\citenamefont {Rau}\ \emph {et~al.}(2014{\natexlab{b}})\citenamefont {Rau}, \citenamefont {Lee},\ and\ \citenamefont {Kee}}]{Rau2014}%
  \BibitemOpen
  \bibfield  {author} {\bibinfo {author} {\bibfnamefont {J.~G.}\ \bibnamefont {Rau}}, \bibinfo {author} {\bibfnamefont {E.~K.-H.}\ \bibnamefont {Lee}},\ and\ \bibinfo {author} {\bibfnamefont {H.-Y.}\ \bibnamefont {Kee}},\ }\href {https://doi.org/10.1103/PhysRevLett.112.077204} {\bibfield  {journal} {\bibinfo  {journal} {Phys. Rev. Lett.}\ }\textbf {\bibinfo {volume} {112}},\ \bibinfo {pages} {077204} (\bibinfo {year} {2014}{\natexlab{b}})}\BibitemShut {NoStop}%
\bibitem [{\citenamefont {Knolle}\ \emph {et~al.}(2018)\citenamefont {Knolle}, \citenamefont {Bhattacharjee},\ and\ \citenamefont {Moessner}}]{Knolle2018}%
  \BibitemOpen
  \bibfield  {author} {\bibinfo {author} {\bibfnamefont {J.}~\bibnamefont {Knolle}}, \bibinfo {author} {\bibfnamefont {S.}~\bibnamefont {Bhattacharjee}},\ and\ \bibinfo {author} {\bibfnamefont {R.}~\bibnamefont {Moessner}},\ }\href {https://doi.org/10.1103/PhysRevB.97.134432} {\bibfield  {journal} {\bibinfo  {journal} {Phys. Rev. B}\ }\textbf {\bibinfo {volume} {97}},\ \bibinfo {pages} {134432} (\bibinfo {year} {2018})}\BibitemShut {NoStop}%
\bibitem [{\citenamefont {Wang}\ \emph {et~al.}(2019)\citenamefont {Wang}, \citenamefont {Normand},\ and\ \citenamefont {Liu}}]{Wang2019}%
  \BibitemOpen
  \bibfield  {author} {\bibinfo {author} {\bibfnamefont {J.}~\bibnamefont {Wang}}, \bibinfo {author} {\bibfnamefont {B.}~\bibnamefont {Normand}},\ and\ \bibinfo {author} {\bibfnamefont {Z.-X.}\ \bibnamefont {Liu}},\ }\href {https://doi.org/10.1103/PhysRevLett.123.197201} {\bibfield  {journal} {\bibinfo  {journal} {Phys. Rev. Lett.}\ }\textbf {\bibinfo {volume} {123}},\ \bibinfo {pages} {197201} (\bibinfo {year} {2019})}\BibitemShut {NoStop}%
\bibitem [{\citenamefont {Gordon}\ \emph {et~al.}(2019)\citenamefont {Gordon}, \citenamefont {Catuneanu}, \citenamefont {S{\o}rensen},\ and\ \citenamefont {Kee}}]{Gordon2019}%
  \BibitemOpen
  \bibfield  {author} {\bibinfo {author} {\bibfnamefont {J.~S.}\ \bibnamefont {Gordon}}, \bibinfo {author} {\bibfnamefont {A.}~\bibnamefont {Catuneanu}}, \bibinfo {author} {\bibfnamefont {E.~S.}\ \bibnamefont {S{\o}rensen}},\ and\ \bibinfo {author} {\bibfnamefont {H.-Y.}\ \bibnamefont {Kee}},\ }\href {https://doi.org/10.1038/s41467-019-10405-8} {\bibfield  {journal} {\bibinfo  {journal} {Nature Communications}\ }\textbf {\bibinfo {volume} {10}},\ \bibinfo {pages} {2470} (\bibinfo {year} {2019})}\BibitemShut {NoStop}%
\bibitem [{\citenamefont {Hickey}\ and\ \citenamefont {Trebst}(2019)}]{Hickey2019}%
  \BibitemOpen
  \bibfield  {author} {\bibinfo {author} {\bibfnamefont {C.}~\bibnamefont {Hickey}}\ and\ \bibinfo {author} {\bibfnamefont {S.}~\bibnamefont {Trebst}},\ }\href {https://doi.org/10.1038/s41467-019-08459-9} {\bibfield  {journal} {\bibinfo  {journal} {Nature Communications}\ }\textbf {\bibinfo {volume} {10}},\ \bibinfo {pages} {530} (\bibinfo {year} {2019})}\BibitemShut {NoStop}%
\bibitem [{\citenamefont {Hwang}\ \emph {et~al.}(2022)\citenamefont {Hwang}, \citenamefont {Go}, \citenamefont {Seong}, \citenamefont {Shibauchi},\ and\ \citenamefont {Moon}}]{Hwang2022}%
  \BibitemOpen
  \bibfield  {author} {\bibinfo {author} {\bibfnamefont {K.}~\bibnamefont {Hwang}}, \bibinfo {author} {\bibfnamefont {A.}~\bibnamefont {Go}}, \bibinfo {author} {\bibfnamefont {J.~H.}\ \bibnamefont {Seong}}, \bibinfo {author} {\bibfnamefont {T.}~\bibnamefont {Shibauchi}},\ and\ \bibinfo {author} {\bibfnamefont {E.-G.}\ \bibnamefont {Moon}},\ }\href {https://doi.org/10.1038/s41467-021-27943-9} {\bibfield  {journal} {\bibinfo  {journal} {Nature Communications}\ }\textbf {\bibinfo {volume} {13}},\ \bibinfo {pages} {323} (\bibinfo {year} {2022})}\BibitemShut {NoStop}%
\bibitem [{\citenamefont {Song}\ \emph {et~al.}(2016)\citenamefont {Song}, \citenamefont {You},\ and\ \citenamefont {Balents}}]{Song2016}%
  \BibitemOpen
  \bibfield  {author} {\bibinfo {author} {\bibfnamefont {X.-Y.}\ \bibnamefont {Song}}, \bibinfo {author} {\bibfnamefont {Y.-Z.}\ \bibnamefont {You}},\ and\ \bibinfo {author} {\bibfnamefont {L.}~\bibnamefont {Balents}},\ }\href {https://doi.org/10.1103/PhysRevLett.117.037209} {\bibfield  {journal} {\bibinfo  {journal} {Phys. Rev. Lett.}\ }\textbf {\bibinfo {volume} {117}},\ \bibinfo {pages} {037209} (\bibinfo {year} {2016})}\BibitemShut {NoStop}%
\bibitem [{\citenamefont {Gotfryd}\ \emph {et~al.}(2017)\citenamefont {Gotfryd}, \citenamefont {Rusna\ifmmode~\check{c}\else \v{c}\fi{}ko}, \citenamefont {Wohlfeld}, \citenamefont {Jackeli}, \citenamefont {Chaloupka},\ and\ \citenamefont {Ole\ifmmode~\acute{s}\else \'{s}\fi{}}}]{Gotfryd2017}%
  \BibitemOpen
  \bibfield  {author} {\bibinfo {author} {\bibfnamefont {D.}~\bibnamefont {Gotfryd}}, \bibinfo {author} {\bibfnamefont {J.}~\bibnamefont {Rusna\ifmmode~\check{c}\else \v{c}\fi{}ko}}, \bibinfo {author} {\bibfnamefont {K.}~\bibnamefont {Wohlfeld}}, \bibinfo {author} {\bibfnamefont {G.}~\bibnamefont {Jackeli}}, \bibinfo {author} {\bibfnamefont {J.~c.~v.}\ \bibnamefont {Chaloupka}},\ and\ \bibinfo {author} {\bibfnamefont {A.~M.}\ \bibnamefont {Ole\ifmmode~\acute{s}\else \'{s}\fi{}}},\ }\href {https://doi.org/10.1103/PhysRevB.95.024426} {\bibfield  {journal} {\bibinfo  {journal} {Phys. Rev. B}\ }\textbf {\bibinfo {volume} {95}},\ \bibinfo {pages} {024426} (\bibinfo {year} {2017})}\BibitemShut {NoStop}%
\bibitem [{\citenamefont {Gohlke}\ \emph {et~al.}(2017)\citenamefont {Gohlke}, \citenamefont {Verresen}, \citenamefont {Moessner},\ and\ \citenamefont {Pollmann}}]{Gohlke2017}%
  \BibitemOpen
  \bibfield  {author} {\bibinfo {author} {\bibfnamefont {M.}~\bibnamefont {Gohlke}}, \bibinfo {author} {\bibfnamefont {R.}~\bibnamefont {Verresen}}, \bibinfo {author} {\bibfnamefont {R.}~\bibnamefont {Moessner}},\ and\ \bibinfo {author} {\bibfnamefont {F.}~\bibnamefont {Pollmann}},\ }\href {https://doi.org/10.1103/PhysRevLett.119.157203} {\bibfield  {journal} {\bibinfo  {journal} {Phys. Rev. Lett.}\ }\textbf {\bibinfo {volume} {119}},\ \bibinfo {pages} {157203} (\bibinfo {year} {2017})}\BibitemShut {NoStop}%
\bibitem [{\citenamefont {Zhang}\ \emph {et~al.}(2021)\citenamefont {Zhang}, \citenamefont {Hal\'asz}, \citenamefont {Zhu},\ and\ \citenamefont {Batista}}]{Zhang2021}%
  \BibitemOpen
  \bibfield  {author} {\bibinfo {author} {\bibfnamefont {S.-S.}\ \bibnamefont {Zhang}}, \bibinfo {author} {\bibfnamefont {G.~B.}\ \bibnamefont {Hal\'asz}}, \bibinfo {author} {\bibfnamefont {W.}~\bibnamefont {Zhu}},\ and\ \bibinfo {author} {\bibfnamefont {C.~D.}\ \bibnamefont {Batista}},\ }\href {https://doi.org/10.1103/PhysRevB.104.014411} {\bibfield  {journal} {\bibinfo  {journal} {Phys. Rev. B}\ }\textbf {\bibinfo {volume} {104}},\ \bibinfo {pages} {014411} (\bibinfo {year} {2021})}\BibitemShut {NoStop}%
\bibitem [{\citenamefont {Willans}\ \emph {et~al.}(2010)\citenamefont {Willans}, \citenamefont {Chalker},\ and\ \citenamefont {Moessner}}]{Willans2010}%
  \BibitemOpen
  \bibfield  {author} {\bibinfo {author} {\bibfnamefont {A.~J.}\ \bibnamefont {Willans}}, \bibinfo {author} {\bibfnamefont {J.~T.}\ \bibnamefont {Chalker}},\ and\ \bibinfo {author} {\bibfnamefont {R.}~\bibnamefont {Moessner}},\ }\href {https://doi.org/10.1103/PhysRevLett.104.237203} {\bibfield  {journal} {\bibinfo  {journal} {Phys. Rev. Lett.}\ }\textbf {\bibinfo {volume} {104}},\ \bibinfo {pages} {237203} (\bibinfo {year} {2010})}\BibitemShut {NoStop}%
\bibitem [{\citenamefont {Willans}\ \emph {et~al.}(2011)\citenamefont {Willans}, \citenamefont {Chalker},\ and\ \citenamefont {Moessner}}]{Willans2011}%
  \BibitemOpen
  \bibfield  {author} {\bibinfo {author} {\bibfnamefont {A.~J.}\ \bibnamefont {Willans}}, \bibinfo {author} {\bibfnamefont {J.~T.}\ \bibnamefont {Chalker}},\ and\ \bibinfo {author} {\bibfnamefont {R.}~\bibnamefont {Moessner}},\ }\href {https://doi.org/10.1103/PhysRevB.84.115146} {\bibfield  {journal} {\bibinfo  {journal} {Phys. Rev. B}\ }\textbf {\bibinfo {volume} {84}},\ \bibinfo {pages} {115146} (\bibinfo {year} {2011})}\BibitemShut {NoStop}%
\bibitem [{\citenamefont {Sreejith}\ \emph {et~al.}(2016)\citenamefont {Sreejith}, \citenamefont {Bhattacharjee},\ and\ \citenamefont {Moessner}}]{Sreejith2016}%
  \BibitemOpen
  \bibfield  {author} {\bibinfo {author} {\bibfnamefont {G.~J.}\ \bibnamefont {Sreejith}}, \bibinfo {author} {\bibfnamefont {S.}~\bibnamefont {Bhattacharjee}},\ and\ \bibinfo {author} {\bibfnamefont {R.}~\bibnamefont {Moessner}},\ }\href {https://doi.org/10.1103/PhysRevB.93.064433} {\bibfield  {journal} {\bibinfo  {journal} {Phys. Rev. B}\ }\textbf {\bibinfo {volume} {93}},\ \bibinfo {pages} {064433} (\bibinfo {year} {2016})}\BibitemShut {NoStop}%
\bibitem [{\citenamefont {Nasu}\ and\ \citenamefont {Motome}(2020)}]{Nasu2020}%
  \BibitemOpen
  \bibfield  {author} {\bibinfo {author} {\bibfnamefont {J.}~\bibnamefont {Nasu}}\ and\ \bibinfo {author} {\bibfnamefont {Y.}~\bibnamefont {Motome}},\ }\href {https://doi.org/10.1103/PhysRevB.102.054437} {\bibfield  {journal} {\bibinfo  {journal} {Phys. Rev. B}\ }\textbf {\bibinfo {volume} {102}},\ \bibinfo {pages} {054437} (\bibinfo {year} {2020})}\BibitemShut {NoStop}%
\bibitem [{\citenamefont {Kao}\ \emph {et~al.}(2021)\citenamefont {Kao}, \citenamefont {Knolle}, \citenamefont {Hal\'asz}, \citenamefont {Moessner},\ and\ \citenamefont {Perkins}}]{Kao2021}%
  \BibitemOpen
  \bibfield  {author} {\bibinfo {author} {\bibfnamefont {W.-H.}\ \bibnamefont {Kao}}, \bibinfo {author} {\bibfnamefont {J.}~\bibnamefont {Knolle}}, \bibinfo {author} {\bibfnamefont {G.~B.}\ \bibnamefont {Hal\'asz}}, \bibinfo {author} {\bibfnamefont {R.}~\bibnamefont {Moessner}},\ and\ \bibinfo {author} {\bibfnamefont {N.~B.}\ \bibnamefont {Perkins}},\ }\href {https://doi.org/10.1103/PhysRevX.11.011034} {\bibfield  {journal} {\bibinfo  {journal} {Phys. Rev. X}\ }\textbf {\bibinfo {volume} {11}},\ \bibinfo {pages} {011034} (\bibinfo {year} {2021})}\BibitemShut {NoStop}%
\bibitem [{\citenamefont {Singhania}\ \emph {et~al.}(2023)\citenamefont {Singhania}, \citenamefont {van~den Brink},\ and\ \citenamefont {Nishimoto}}]{Singhania2023}%
  \BibitemOpen
  \bibfield  {author} {\bibinfo {author} {\bibfnamefont {A.}~\bibnamefont {Singhania}}, \bibinfo {author} {\bibfnamefont {J.}~\bibnamefont {van~den Brink}},\ and\ \bibinfo {author} {\bibfnamefont {S.}~\bibnamefont {Nishimoto}},\ }\href {https://doi.org/10.1103/PhysRevResearch.5.023009} {\bibfield  {journal} {\bibinfo  {journal} {Phys. Rev. Res.}\ }\textbf {\bibinfo {volume} {5}},\ \bibinfo {pages} {023009} (\bibinfo {year} {2023})}\BibitemShut {NoStop}%
\bibitem [{\citenamefont {Zschocke}\ and\ \citenamefont {Vojta}(2015)}]{Zschocke20152}%
  \BibitemOpen
  \bibfield  {author} {\bibinfo {author} {\bibfnamefont {F.}~\bibnamefont {Zschocke}}\ and\ \bibinfo {author} {\bibfnamefont {M.}~\bibnamefont {Vojta}},\ }\href {https://doi.org/10.1103/PhysRevB.92.014403} {\bibfield  {journal} {\bibinfo  {journal} {Phys. Rev. B}\ }\textbf {\bibinfo {volume} {92}},\ \bibinfo {pages} {014403} (\bibinfo {year} {2015})}\BibitemShut {NoStop}%
\bibitem [{\citenamefont {Dhochak}\ \emph {et~al.}(2010)\citenamefont {Dhochak}, \citenamefont {Shankar},\ and\ \citenamefont {Tripathi}}]{Dhochak2010}%
  \BibitemOpen
  \bibfield  {author} {\bibinfo {author} {\bibfnamefont {K.}~\bibnamefont {Dhochak}}, \bibinfo {author} {\bibfnamefont {R.}~\bibnamefont {Shankar}},\ and\ \bibinfo {author} {\bibfnamefont {V.}~\bibnamefont {Tripathi}},\ }\href {https://doi.org/10.1103/PhysRevLett.105.117201} {\bibfield  {journal} {\bibinfo  {journal} {Phys. Rev. Lett.}\ }\textbf {\bibinfo {volume} {105}},\ \bibinfo {pages} {117201} (\bibinfo {year} {2010})}\BibitemShut {NoStop}%
\bibitem [{\citenamefont {Kimchi}\ \emph {et~al.}(2018)\citenamefont {Kimchi}, \citenamefont {Nahum},\ and\ \citenamefont {Senthil}}]{Kimchi2018}%
  \BibitemOpen
  \bibfield  {author} {\bibinfo {author} {\bibfnamefont {I.}~\bibnamefont {Kimchi}}, \bibinfo {author} {\bibfnamefont {A.}~\bibnamefont {Nahum}},\ and\ \bibinfo {author} {\bibfnamefont {T.}~\bibnamefont {Senthil}},\ }\href {https://doi.org/10.1103/PhysRevX.8.031028} {\bibfield  {journal} {\bibinfo  {journal} {Phys. Rev. X}\ }\textbf {\bibinfo {volume} {8}},\ \bibinfo {pages} {031028} (\bibinfo {year} {2018})}\BibitemShut {NoStop}%
\bibitem [{\citenamefont {Pereira}\ \emph {et~al.}(2006)\citenamefont {Pereira}, \citenamefont {Guinea}, \citenamefont {Lopes~dos Santos}, \citenamefont {Peres},\ and\ \citenamefont {Castro~Neto}}]{Pereira2006}%
  \BibitemOpen
  \bibfield  {author} {\bibinfo {author} {\bibfnamefont {V.~M.}\ \bibnamefont {Pereira}}, \bibinfo {author} {\bibfnamefont {F.}~\bibnamefont {Guinea}}, \bibinfo {author} {\bibfnamefont {J.~M.~B.}\ \bibnamefont {Lopes~dos Santos}}, \bibinfo {author} {\bibfnamefont {N.~M.~R.}\ \bibnamefont {Peres}},\ and\ \bibinfo {author} {\bibfnamefont {A.~H.}\ \bibnamefont {Castro~Neto}},\ }\href {https://doi.org/10.1103/PhysRevLett.96.036801} {\bibfield  {journal} {\bibinfo  {journal} {Phys. Rev. Lett.}\ }\textbf {\bibinfo {volume} {96}},\ \bibinfo {pages} {036801} (\bibinfo {year} {2006})}\BibitemShut {NoStop}%
\bibitem [{sup()}]{supplementaryMaterial}%
  \BibitemOpen
  \href@noop {} {}\bibinfo {howpublished} {See Supplementary Material for the correlation functions at varying distances, the effect of bond disorder, an illustration of the interaction between two phantom spins, a discussion on other integrable models, and the effective coupling scalings for popular microscopic interactions, which includes \refscite{Pereira2008,Sanyal2021}}\BibitemShut {NoStop}%
\bibitem [{\citenamefont {Vojta}\ \emph {et~al.}(2016)\citenamefont {Vojta}, \citenamefont {Mitchell},\ and\ \citenamefont {Zschocke}}]{PhysRevLett.117.037202}%
  \BibitemOpen
  \bibfield  {author} {\bibinfo {author} {\bibfnamefont {M.}~\bibnamefont {Vojta}}, \bibinfo {author} {\bibfnamefont {A.~K.}\ \bibnamefont {Mitchell}},\ and\ \bibinfo {author} {\bibfnamefont {F.}~\bibnamefont {Zschocke}},\ }\href {https://doi.org/10.1103/PhysRevLett.117.037202} {\bibfield  {journal} {\bibinfo  {journal} {Phys. Rev. Lett.}\ }\textbf {\bibinfo {volume} {117}},\ \bibinfo {pages} {037202} (\bibinfo {year} {2016})}\BibitemShut {NoStop}%
\bibitem [{\citenamefont {Das}\ \emph {et~al.}(2016)\citenamefont {Das}, \citenamefont {Dhochak},\ and\ \citenamefont {Tripathi}}]{PhysRevB.94.024411}%
  \BibitemOpen
  \bibfield  {author} {\bibinfo {author} {\bibfnamefont {S.~D.}\ \bibnamefont {Das}}, \bibinfo {author} {\bibfnamefont {K.}~\bibnamefont {Dhochak}},\ and\ \bibinfo {author} {\bibfnamefont {V.}~\bibnamefont {Tripathi}},\ }\href {https://doi.org/10.1103/PhysRevB.94.024411} {\bibfield  {journal} {\bibinfo  {journal} {Phys. Rev. B}\ }\textbf {\bibinfo {volume} {94}},\ \bibinfo {pages} {024411} (\bibinfo {year} {2016})}\BibitemShut {NoStop}%
\bibitem [{\citenamefont {Jayaprakash}\ \emph {et~al.}(1981)\citenamefont {Jayaprakash}, \citenamefont {Krishna-murthy},\ and\ \citenamefont {Wilkins}}]{PhysRevLett.47.737}%
  \BibitemOpen
  \bibfield  {author} {\bibinfo {author} {\bibfnamefont {C.}~\bibnamefont {Jayaprakash}}, \bibinfo {author} {\bibfnamefont {H.~R.}\ \bibnamefont {Krishna-murthy}},\ and\ \bibinfo {author} {\bibfnamefont {J.~W.}\ \bibnamefont {Wilkins}},\ }\href {https://doi.org/10.1103/PhysRevLett.47.737} {\bibfield  {journal} {\bibinfo  {journal} {Phys. Rev. Lett.}\ }\textbf {\bibinfo {volume} {47}},\ \bibinfo {pages} {737} (\bibinfo {year} {1981})}\BibitemShut {NoStop}%
\bibitem [{\citenamefont {Jones}\ and\ \citenamefont {Varma}(1987)}]{PhysRevLett.58.843}%
  \BibitemOpen
  \bibfield  {author} {\bibinfo {author} {\bibfnamefont {B.~A.}\ \bibnamefont {Jones}}\ and\ \bibinfo {author} {\bibfnamefont {C.~M.}\ \bibnamefont {Varma}},\ }\href {https://doi.org/10.1103/PhysRevLett.58.843} {\bibfield  {journal} {\bibinfo  {journal} {Phys. Rev. Lett.}\ }\textbf {\bibinfo {volume} {58}},\ \bibinfo {pages} {843} (\bibinfo {year} {1987})}\BibitemShut {NoStop}%
\bibitem [{\citenamefont {Yao}\ and\ \citenamefont {Kivelson}(2007)}]{Yao2007}%
  \BibitemOpen
  \bibfield  {author} {\bibinfo {author} {\bibfnamefont {H.}~\bibnamefont {Yao}}\ and\ \bibinfo {author} {\bibfnamefont {S.~A.}\ \bibnamefont {Kivelson}},\ }\href {https://doi.org/10.1103/PhysRevLett.99.247203} {\bibfield  {journal} {\bibinfo  {journal} {Phys. Rev. Lett.}\ }\textbf {\bibinfo {volume} {99}},\ \bibinfo {pages} {247203} (\bibinfo {year} {2007})}\BibitemShut {NoStop}%
\bibitem [{\citenamefont {Yao}\ and\ \citenamefont {Lee}(2011)}]{Yao2011}%
  \BibitemOpen
  \bibfield  {author} {\bibinfo {author} {\bibfnamefont {H.}~\bibnamefont {Yao}}\ and\ \bibinfo {author} {\bibfnamefont {D.-H.}\ \bibnamefont {Lee}},\ }\href {https://doi.org/10.1103/PhysRevLett.107.087205} {\bibfield  {journal} {\bibinfo  {journal} {Phys. Rev. Lett.}\ }\textbf {\bibinfo {volume} {107}},\ \bibinfo {pages} {087205} (\bibinfo {year} {2011})}\BibitemShut {NoStop}%
\bibitem [{\citenamefont {Pereira}\ \emph {et~al.}(2008)\citenamefont {Pereira}, \citenamefont {Lopes~dos Santos},\ and\ \citenamefont {Castro~Neto}}]{Pereira2008}%
  \BibitemOpen
  \bibfield  {author} {\bibinfo {author} {\bibfnamefont {V.~M.}\ \bibnamefont {Pereira}}, \bibinfo {author} {\bibfnamefont {J.~M.~B.}\ \bibnamefont {Lopes~dos Santos}},\ and\ \bibinfo {author} {\bibfnamefont {A.~H.}\ \bibnamefont {Castro~Neto}},\ }\href {https://doi.org/10.1103/PhysRevB.77.115109} {\bibfield  {journal} {\bibinfo  {journal} {Phys. Rev. B}\ }\textbf {\bibinfo {volume} {77}},\ \bibinfo {pages} {115109} (\bibinfo {year} {2008})}\BibitemShut {NoStop}%
\bibitem [{\citenamefont {Sanyal}\ \emph {et~al.}(2021)\citenamefont {Sanyal}, \citenamefont {Damle}, \citenamefont {Chalker},\ and\ \citenamefont {Moessner}}]{Sanyal2021}%
  \BibitemOpen
  \bibfield  {author} {\bibinfo {author} {\bibfnamefont {S.}~\bibnamefont {Sanyal}}, \bibinfo {author} {\bibfnamefont {K.}~\bibnamefont {Damle}}, \bibinfo {author} {\bibfnamefont {J.~T.}\ \bibnamefont {Chalker}},\ and\ \bibinfo {author} {\bibfnamefont {R.}~\bibnamefont {Moessner}},\ }\href {https://doi.org/10.1103/PhysRevLett.127.127201} {\bibfield  {journal} {\bibinfo  {journal} {Phys. Rev. Lett.}\ }\textbf {\bibinfo {volume} {127}},\ \bibinfo {pages} {127201} (\bibinfo {year} {2021})}\BibitemShut {NoStop}%
\end{thebibliography}%
\appendix
\newpage
\section*{{Supplementary Material for: Vacancies in generic Kitaev spin liquids}}
\section{The Green function of a honeycomb lattice on an infinite cylinder with defects}
 \begin{figure}[htbp]
   \centering
 \includegraphics[width=0.99\linewidth]{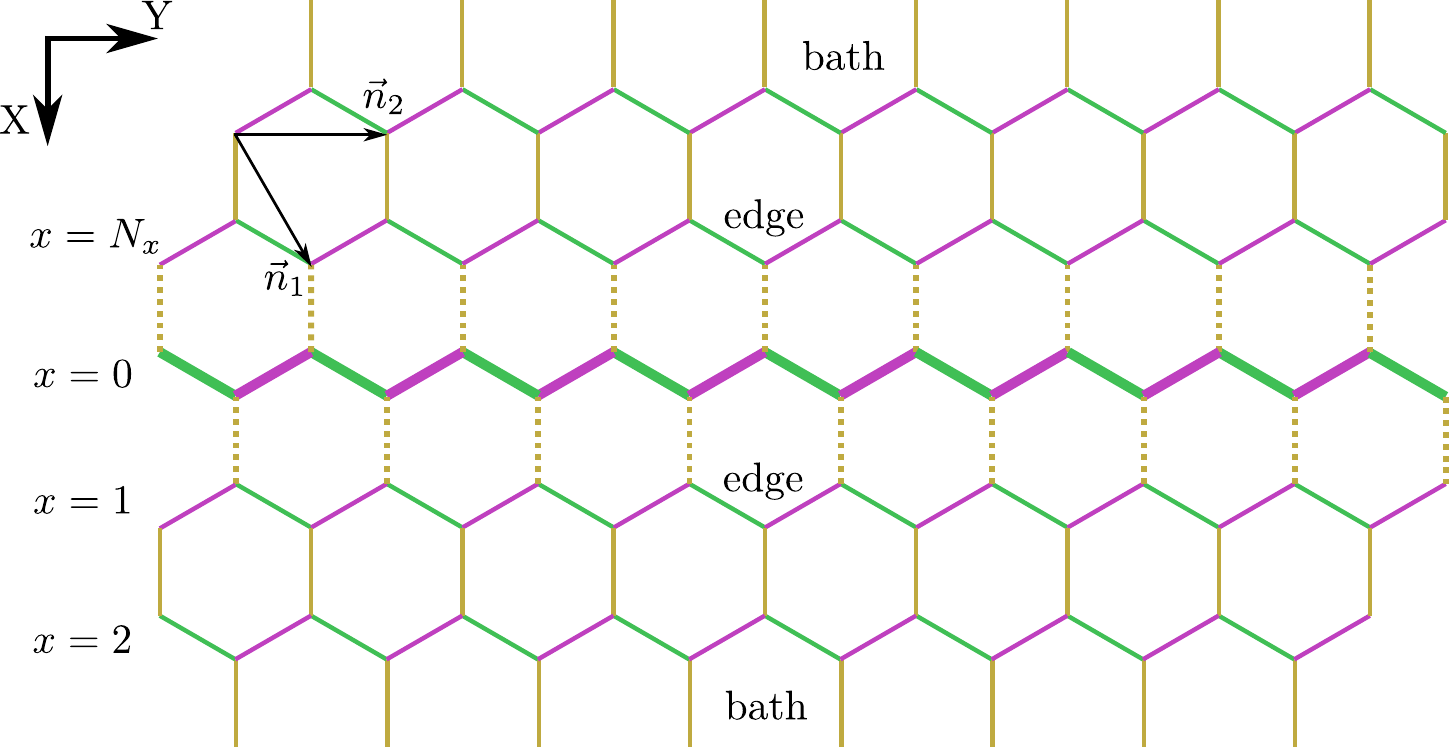}
   \caption{We treat the two-dimensional honeycomb lattice as a one-dimensional chain coupled to a two-dimensional bath. The former contains all $x=0$ sites and is indicated by thick links. Thin links indicate couplings within the bath, and dotted lines are the chain--bath couplings. The separation into chain and bath is purely formal; the values of all nearest-neighbor couplings are identical. We perform our numerical calculations on an infinite cylinder with $\vec n_2$ the periodic direction.}
   \label{fig: new_method_schematics}
 \end{figure}
In the main text, we are interested in the Green functions for a honeycomb lattice with vacancies or flux excitations. To obtain them efficiently, we split the system into a single chain ($x=0$ in Fig.~\ref{fig: new_method_schematics}) and semi-infinite baths on either side of it. The former contains all defects; the latter are translation invariant in the periodic direction. The Green function on the central region is then given by
\begin{align}
G(i\omega) = [i\omega - H^\text{center}-\Sigma(i\omega)]^{-1} \label{eqn:greencyl},
\end{align}
where $ H^\text{center}$ is the Hamiltonian for the central cylinder. The self-energy $\Sigma(i\omega)$ is determined by the Green function $G_\text{bath}(i\omega)$ of the semi-infinite bath via
\begin{align}
\Sigma(i\omega) = H^\text{center-bath} G_\text{bath}(i\omega) H^\text{bath-center} ~,
\label{eqn:selfenergy}\end{align}
where $H^\text{center-bath}$ contains any elements that connect the central cylinder to the bath. Crucially, an analytical expression for the relevant elements of $G_\text{bath}$ can be easily obtained for any momentum $k_y$ in the periodic direction. Inserting the thus-obtained $\Sigma(i\omega)$ into Eq.~\eqref{eqn:greencyl} yields the full Green function $G(i\omega)$ of the central region. In essence, the calculation of $G(i\omega)$ reduces to solving an effective quasi-one-dimensional model.

To obtain the bath Green function, we begin with an infinite cylinder that is translation invariant in both directions. Its Green function $G^0$ is given by
\begin{align}
    &G^0_{\vect k}(i\omega) =-\frac{1}{\omega^2+|g(\vect k)|^2} \begin{pmatrix} i\omega & ig(\vect k)\\ -ig^*(\vect k) & i\omega \end{pmatrix}, \\
    & g(\vect k)=1+e^{ik_x}+e^{ik_y}.
\end{align}
To obtain the semi-infinite bath, we add a potential barrier $V$ for all $x =0$ (say) and compute the new Green function from $G =[G^0+V]^{-1}$. Taking $V\rightarrow \infty$, we thus obtain
\begin{align}
G^\text{bath}_{x,x'}(k_y) = 
G^0_{x,x'}(k_y)-
\frac{G^0_{x,0}(k_y) G^0_{0,x'}(k_y)}{G^0_{0,0}(k_y)}~.
\end{align}
For Eq.~\ref{eqn:selfenergy} with $H^\text{center-bath}$, which only contains nearest-neighbor hopping, we only need $G^\text{bath}_{1,1}(k_y)$, for which we find
\begin{align}
&G^\text{bath}_{1,1}(k_y)=i\left(\frac{1+\omega^2+2\cos(k_y)}{2 \omega}\right)\\
    &-i\left(\frac{ \sqrt{8[\omega^2+\cos(k_y)]\cos^2(k_y/2)+(1+\omega^2)^2}}{2 \omega}\right).\nonumber
    \label{eq: GF_analytic_formula}
\end{align}

\section{Two vacancies on a honeycomb lattice}
\label{appb}
It is well-known that vacancies on the same sublattice provide independent zero modes, while zero modes of the vacancies on opposite sublattices can hybridize \cite{Pereira2008}. The strength of this hybridization, and thus the degree to which the degeneracy is lifted, depends sensitively on the relative location of the two vacancies. To understand this property, we compute the matrix element
\begin{align}
M_{\vect R_A,\vect R_B} \equiv \sum_{\vect r,\vect r'}
\Psi^*_{\vect R_A}(\vect r) H_{\vect R_A,\vect R_B}(\vect r,\vect r')
\Psi_{\vect R_B}(\vect r ')~,
\end{align}
where $\Psi_{\vect R}(\vect r ')$ is the zero-mode wave function for a honeycomb lattice with a  \text{single} vacancy and  $H_{\vect R_A,\vect R_B}$ is the Hamiltonian of the same system with two vacancies. Since $\Psi_{\vect R}(\vect r ')$ is an exact zero-mode of the single-vacancy model
\begin{align}
H_{\vect R_B}&=
H_{\vect R_A,\vect R_B}- iJ \sum_{\mu=x,y,z}|\vect R_A\rangle\langle  \vect R_A + \hat e_\mu|\\
&+iJ \sum_{\mu=x,y,z}|\vect R_A + \hat e_\mu \rangle \langle \vect R_A |~,
\end{align}
we can simplify
\begin{align}
M_{\vect R_A,\vect R_B} =-
 iJ \sum_{\mu=x,y,z} \Psi^*_{\vect R_A}( \vect R_A +\hat e_\mu )
\Psi_{\vect R_B}(\vect R_A) ~.
\end{align}
To understand the behavior of $M_{\vect R_A,\vect R_B}$, it is thus sufficient to look at $\Psi_{\vect R_B}(\vect R_A )$, which was computed in Ref.~\cite{Pereira2006}. For $\vect R_A =\vect R_B - \hat e_y - n(\hat e_y-\hat e_x)$, we find
\begin{align}
\lim\limits_{n\rightarrow \infty}
|\Psi_{\vect R_B}(\vect R_A ) | \sim \begin{cases}
\frac{1}{n} \quad  &n = 0\text{ mod } 3~,
\\
\frac{1}{n} \quad  &n = 2\text{ mod } 3~,
\\
\frac{1}{n^2} \quad  &n = 1\text{ mod } 3~.
\end{cases}
\end{align}
For the first two cases, the smallest hybridization in a finite system is parameterically the same as the finite-size splitting for the bulk modes. As such, they do not substantially modify the low-energy Green function. By contrast, in the third case, the energy of the nearly-zero modes $(\Psi_{\vect R_A}\pm i\Psi_{\vect R_B})/\sqrt{2}$ is parameterically smaller and will dominate the low-energy Green function. As this property relies on a careful cancellation between different $1/n$ contributions, we deem it less generic and focus on the other cases in the main text. 

\subsection{Anomalous low-energy modes for specific vacancy location}

To confirm the behavior derived above, we numerically computed the spectrum for systems of size $N_x=N_y = 8+6k+1$ and $N_x=N_y = 6+6k+1, k \in \mathbb{Z}$, which support periodic $n = 1\text{ mod } 3$ and $n = 0\text{ mod } 3$ subsequences along the $y$ direction, respectively. We place two vacancies along the chain $x = \text{const}$ and separate them by $N_y/2$ unit cells such that the distance between the vacancy sites belongs to the corresponding subsequence. 

\begin{figure}
   \centering
 \includegraphics[width=0.99\linewidth]{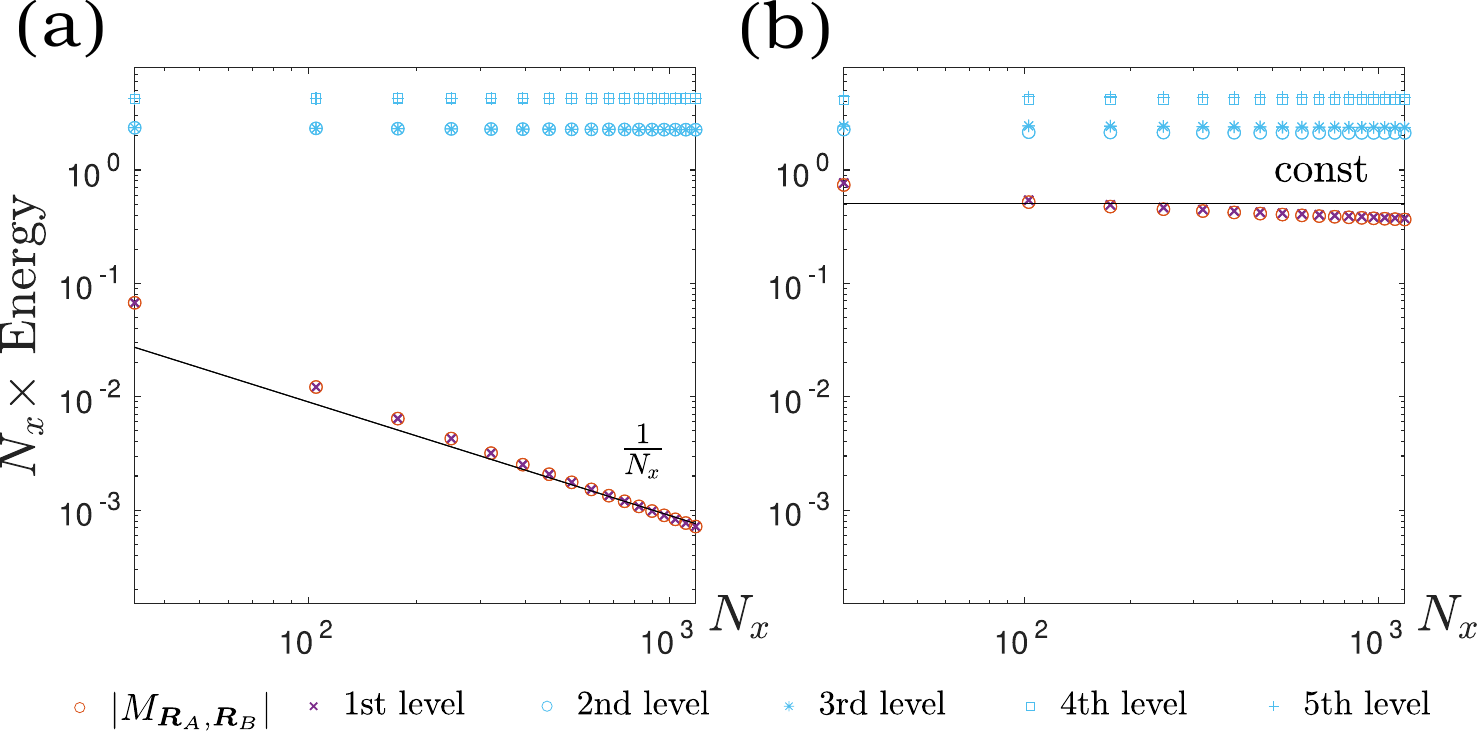}
   \caption{The energies of the first five energy levels as functions of the system size when the distance between the vacancies is approximately $N_y/2$. The columns (a) and (b) correspond to the $n = 1\text{ mod } 3$ and $n = 0\text{ mod } 3$ distances, respectively. The orange circles show the amplitude $M_{\vect R_A,\vect R_B}$. The power laws with definite powers (the solid black lines) are put as reference lines to illustrate the lowest energy state dependence.}
   \label{fig: 2_vacs_good_bad_subs_app}
 \end{figure}

For $n = 1\text{ mod } 3$, the lowest energy state occurs well separated from all other states; its energy decays parametrically differently with vacancy separation (see \figref{fig: 2_vacs_good_bad_subs_app}).
The lowest energy state agrees with the matrix element $|M_{\vect R_A,\vect R_B}|$ within a relative error of less than a few percent when $|\vec{r}_{v1}-\vec{r}_{v2}| \gg 1$, which indicates the zero mode origin of that state. We note that the lowest energy state in either case does not strictly follow the suggested power law. These deviations arise due to the zero-mode normalization factors, which decay logarithmically with system size.

\begin{figure}
   \centering
 \includegraphics[width=0.99\linewidth]{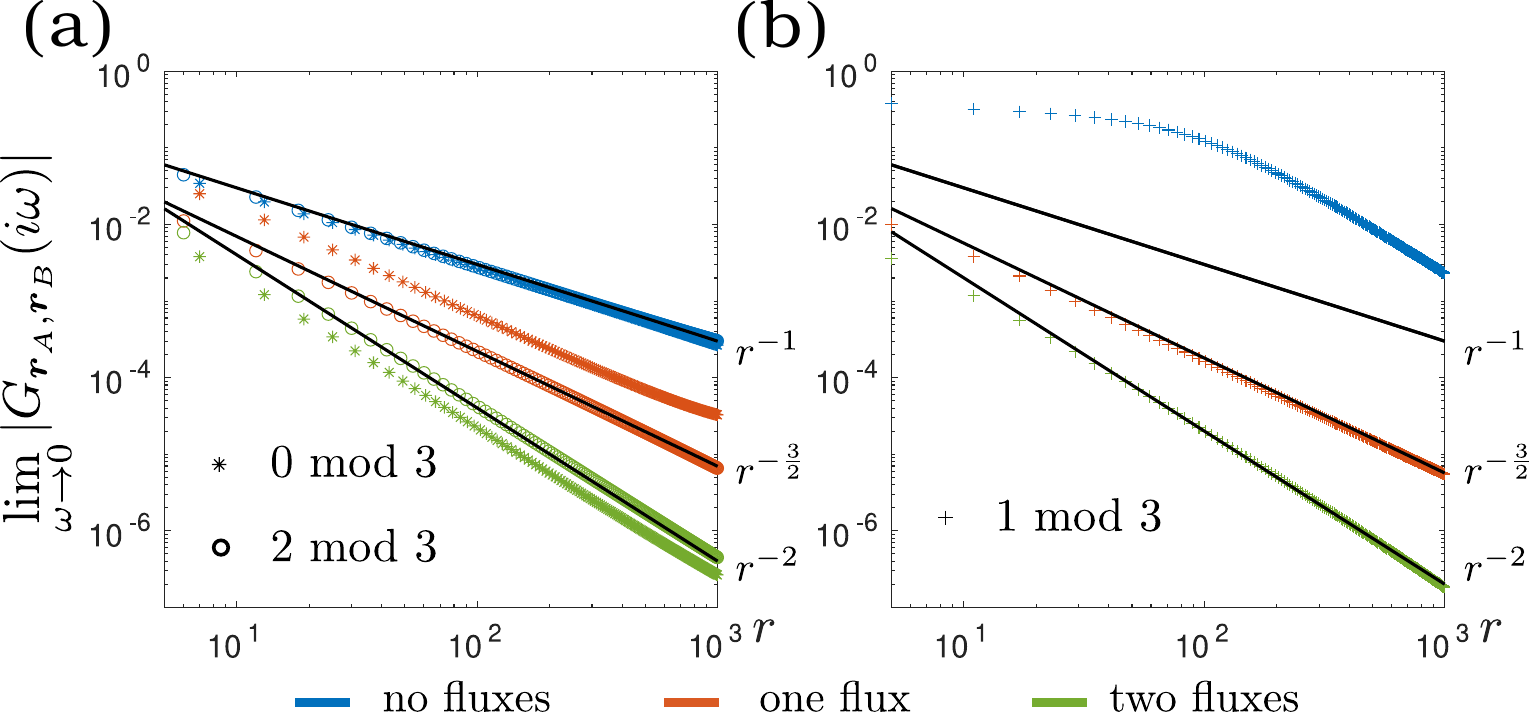}
   \caption{Green functions between the sites adjacent to two vacancies separated along the periodic direction of an infinite cylinder with a $10000$ unit cell circumference. For vacancy separations given by  $n=0$ mod $3$ or $n=2$ mod $3$ unit cells, we find clear power laws that depend on the presence of fluxes at the vacancies (a). For separations $n = 1$ mod $3$, the Green functions follow a qualitatively different decay when neither vacancy traps flux (b). When either of them traps a flux, we recover the same power-law decay as in (a). }
   \label{fig: 2_vac_green_func_app}
 \end{figure}

In the presence of anomalous low-energy modes, the  Green functions change qualitatively. \figref{fig: 2_vac_green_func_app} (a) and (b) show the Green function without and with such modes, respectively. In the latter case, it does not follow a clear power law. When either vacancy traps a flux, there are no anomalous low-energy states for any vacancy separation. In those cases, the Green functions decay with identical power laws for all sequences.

 \subsection{Correlation functions with another pair of vacancies}
  \begin{figure}
   \centering
 \includegraphics[width=0.99\linewidth]{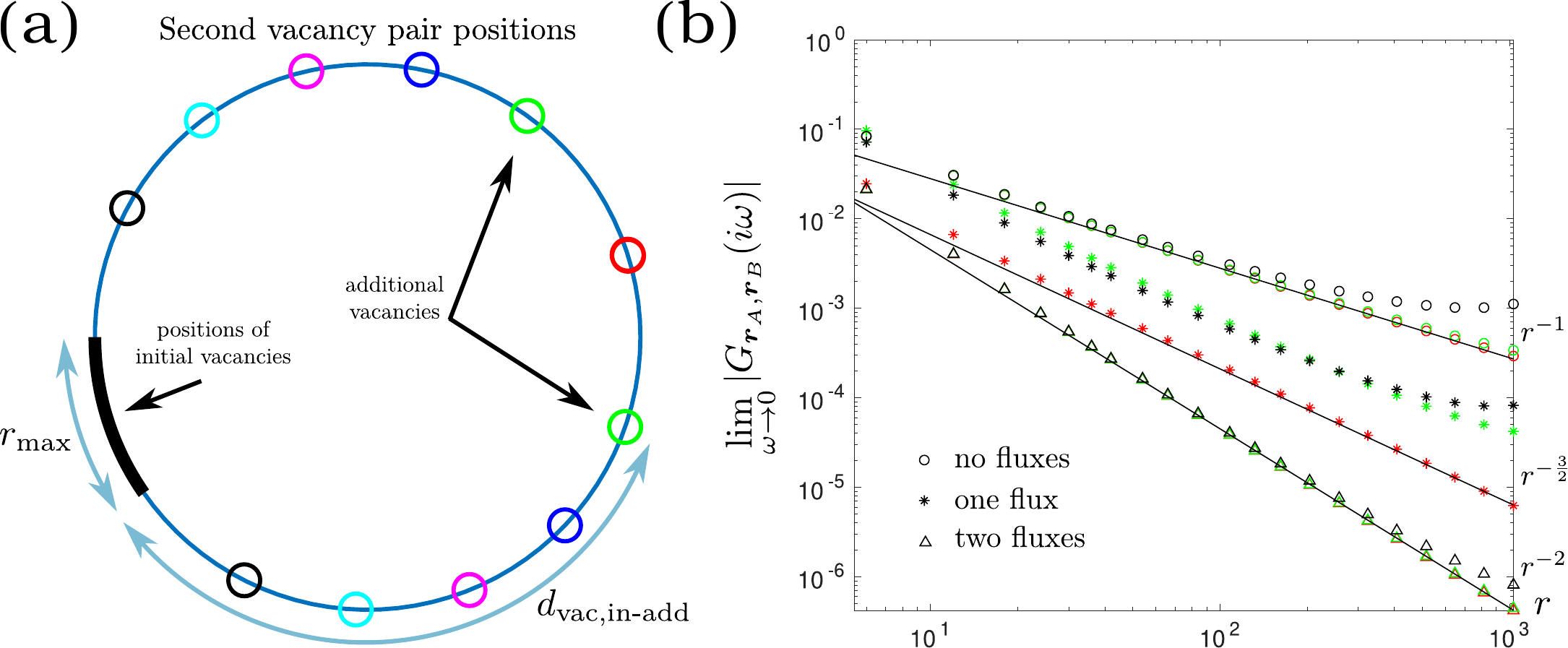}
   \caption{(a) Schematics of the positions of four vacancies. The large blue circle represents a periodic chain with four vacancies. The thick black line corresponds to the positions of the initial pair of vacancies: the first vacancy is fixed at the top of a black segment, and the second is moved down along the segment. The pairs of colored circles illustrate the additional vacancy pair whose positions are fixed when computing the correlation function between the initial vacancies. (b) The correlation function for three positions of additional vacancies and a fixed distance subsequence. Different colors correspond to different positions in (a).}
   \label{fig: second_pair_green_func}
 \end{figure}
 In this subsection, we show how correlations change in the presence of two additional vacancies along a chain (see Fig. \ref{fig: second_pair_green_func} (a)).  We fixed their position and computed the correlation function between the initial pair of vacancies for one of the distance subsequences. The correlation functions for additional vacancy separations $d_{\text{vac,in-add}} \approx r_{\text{max}}, 5 r_{\text{max}}, 6 r_{\text{max}}$ are shown in Fig. \ref{fig: second_pair_green_func} (b). The presence of the other two vacancies barely affects the power law estimates. For a single flux attached to a vacancy (and the other one sitting at $N_y/2$), different positions of additional vacancies also lead to a change of the subsequence. This is a consequence of a second flux affecting the system. It is also observed without additional vacancies when the second flux position is chosen differently from $N_y/2$.  In all cases, deviations from the predicted power-law decays become noticeable near $r = r_{\text{max}}$; however, different flux configurations still exhibit differences of several orders of magnitude.

\section{Delocalized zero modes in the presence of bond disorder} 
\begin{figure}[h]
   \centering
 \includegraphics[width=0.99\linewidth]{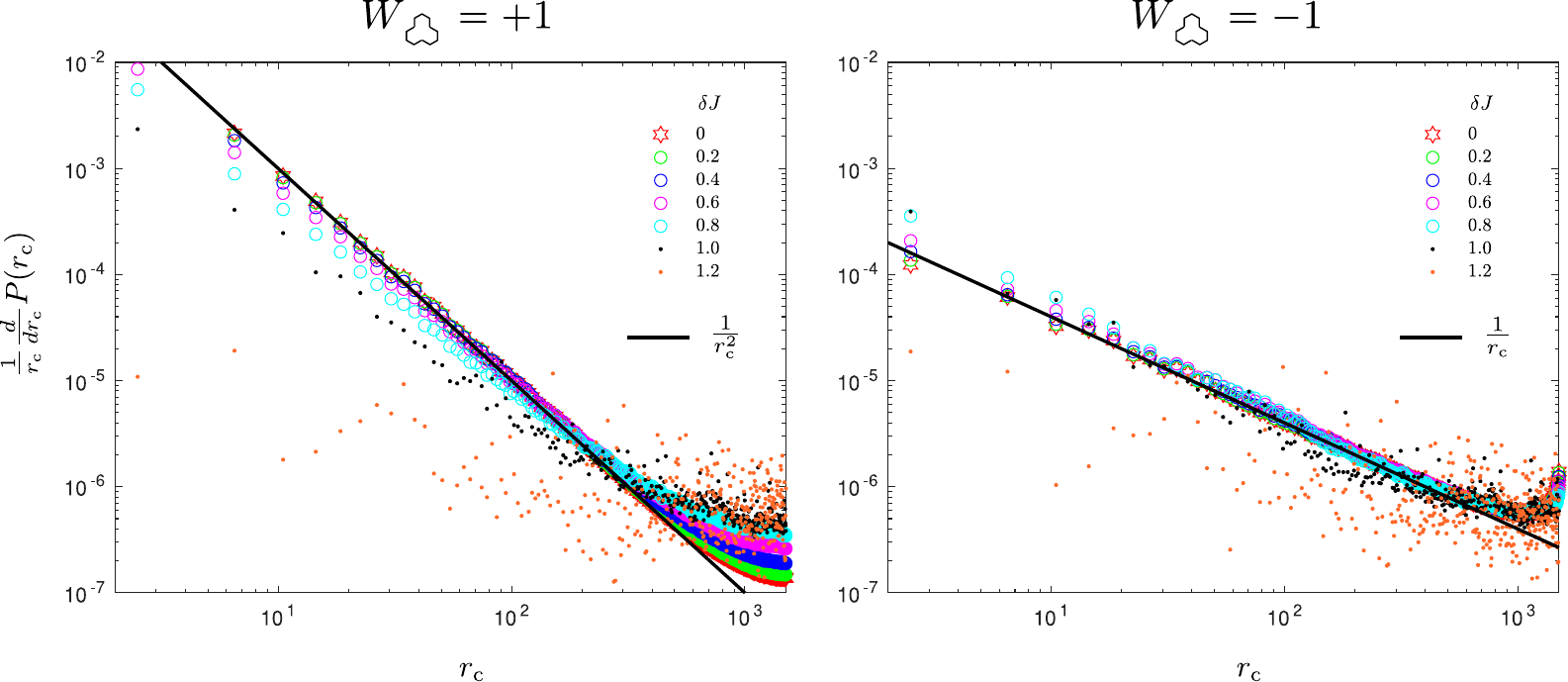}
   \caption{The decay of the average zero mode density shown on the log-log scale. The lattice is periodic, with a size of $3001 \times 3002$ unit cells and a vacancy at the center. Noise is added to the model with $J = -1$. Stochastic averaging is performed over an ensemble of 500 runs. In the absence of flux, the density decays as $1/r_{\text{c}}^2$. In the presence of flux, the scaling is $1/r_{\text{c}}$. These scaling properties remain largely unchanged unless the noise magnitude approaches the Kitaev coupling.}
   \label{fig: delocal_zero_modes}
 \end{figure}
 We further investigate the impact of the bond disorder on the vacancy magnetization and stability of power-law decays. In brief, even moderately strong bond disorder does not change the power-law decay of zero modes. We consequently expect that the Green functions will also retain their scaling behavior. However, their direct computation at the system sizes shown in the main text is not possible, as the scattering approach that we use requires a clean bath.
 
 Previous exact diagonalization studies of the KSL with bond disorder found that noise with a width up to $0.5|J|$ keeps the ground state flux-free, but changes the KSL stability region \cite{Singhania2023}. Studies of quadratic models showed that the same disorder magnitude $0.5|J|$ decreases the size of the flux gaps \cite{Zschocke20152} and can exhibit flux excitations. We consider bond disorder by adding uniformly distributed noise to the Kitaev couplings,
\begin{equation}
J \to J+\delta J_{\vect r \vect r'}~, \quad \delta J_{\vect r \vect r'} \in [-\delta J, +\delta J]~.
\end{equation}
We then analyze how this coupling noise affects the localization of the zero modes for two cases: flux-free and vacancy flux. In the latter case, the second flux is at the boundary of the system, far from the vacancy, which is in the system's center. Specifically, we compute the average zero-mode density integrated within a circle 
\begin{equation}
    P(r_{\text{c}}) = \sum\limits_{|\vect r -\vect R| \leq r_{\text{c}}} \langle |\psi^0_{\vect R}(\vect r)|^2 \rangle_{\delta J_{\vect r \vect r'}}~,
\end{equation}
and use it to find the derivative $\frac{1}{r_{\text{c}}} \frac{d}{d r_{\text{c}}} P(r_{\text{c}})$, which gives the average density over a thin ring at distance $r_{\text{c}}$.

We find that mode localizations remain stable for moderately strong disorder $\delta J \geq 0.5|J|$ (see Fig. \ref{fig: delocal_zero_modes}). For even stronger disorder, the ground state is no longer flux-free, and one should additionally optimize the flux configurations. Still, the behavior in that limit can be inferred from $\delta J> |J|$, where some bonds have opposite signs, which introduces additional background fluxes. At this point, we no longer find zero-modes centered on vacancy sites.

\section{Exact diagonalization of a 24-site cluster with two vacancies}
\label{app: ED_app}
 \begin{figure}
   \centering
 \includegraphics[width=0.99\linewidth]{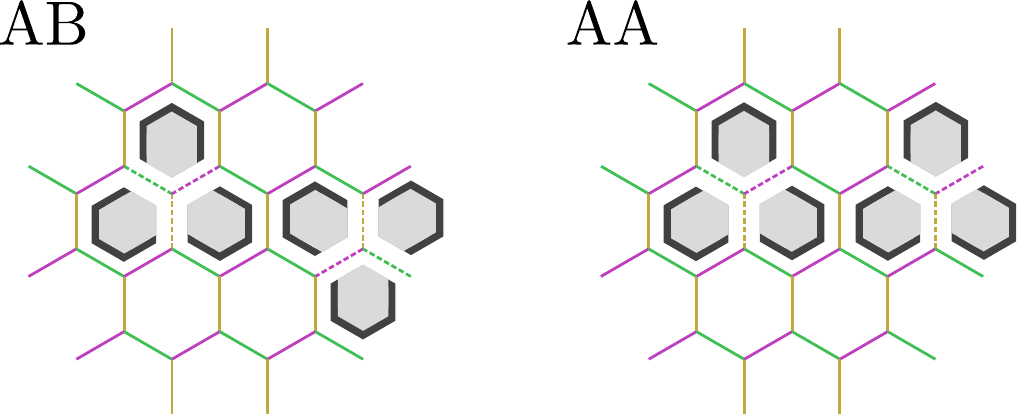}
   \caption{Two vacancies in the 24-site cluster. The dotted lines and five-spin $\tau$ operators indicate the vacancy sites. The external lines of the same color show how the cluster is periodically connected.
   }
\label{fig:two_vacancies_deg_lifting}
 \end{figure}
To verify the predicted lifting of degeneracies, we performed exact diagonalization of a 24-site cluster with two vacancies on equal or opposite sublattices (See Fig.~\ref{fig:two_vacancies_deg_lifting}). We computed energy spectra for
\begin{align}
H_1&=H_\text{K,anis} + H_\text{quad } \\
H_2&=H_\text{K,anis} + H_\text{quad} + H_\text{quart }\\
H_3&=H_\text{K,anis} + H_{\text{Heis.}-\Gamma}
\end{align}
 with
\begin{align}
H_\text{K,anis}&= \sum_{ \vect r ,\mu} J^\mu  \sigma^{\mu}_{\vect r} \sigma^{\mu}_{\vect r + \hat e_\mu} \\
   H_\text{quad }&= \frac{J_{\text{quad}}}{2} \sum_{ \vect R ,\mu \neq \kappa \neq \nu} (1+ \epsilon_{\mu\nu\kappa})  \sigma^{\kappa}_{\vect R +\hat{e}_\mu} \sigma^{\nu}_{\vect R +\hat{e}_\mu + \hat{e}_\nu}  \\
 H_\text{quart}&= J_{\text{quart}} \sum_{ \vect R , \alpha}\sigma_{\vect R +\hat{e}_{\alpha}} \tau^{\alpha}_{\vect R}\\
 H_{\text{Heis.}} &= J_{\text{Heis.}-\Gamma}  \sum_{\langle\vect  r \vect r'\rangle} \vect \sigma_{\vect r} \cdot \vect \sigma_{\vect r'}  \\
 H_{\Gamma} &=  J_{\text{Heis.}-\Gamma} \sum_{\vect r,\mu}|\epsilon_{\mu\nu\kappa}|\sigma_{\vect r}^{\nu} \sigma_{\vect r+\hat e_\mu}^{\kappa}
\end{align}
\begin{figure}
   \centering
 \includegraphics[width=0.99\linewidth]{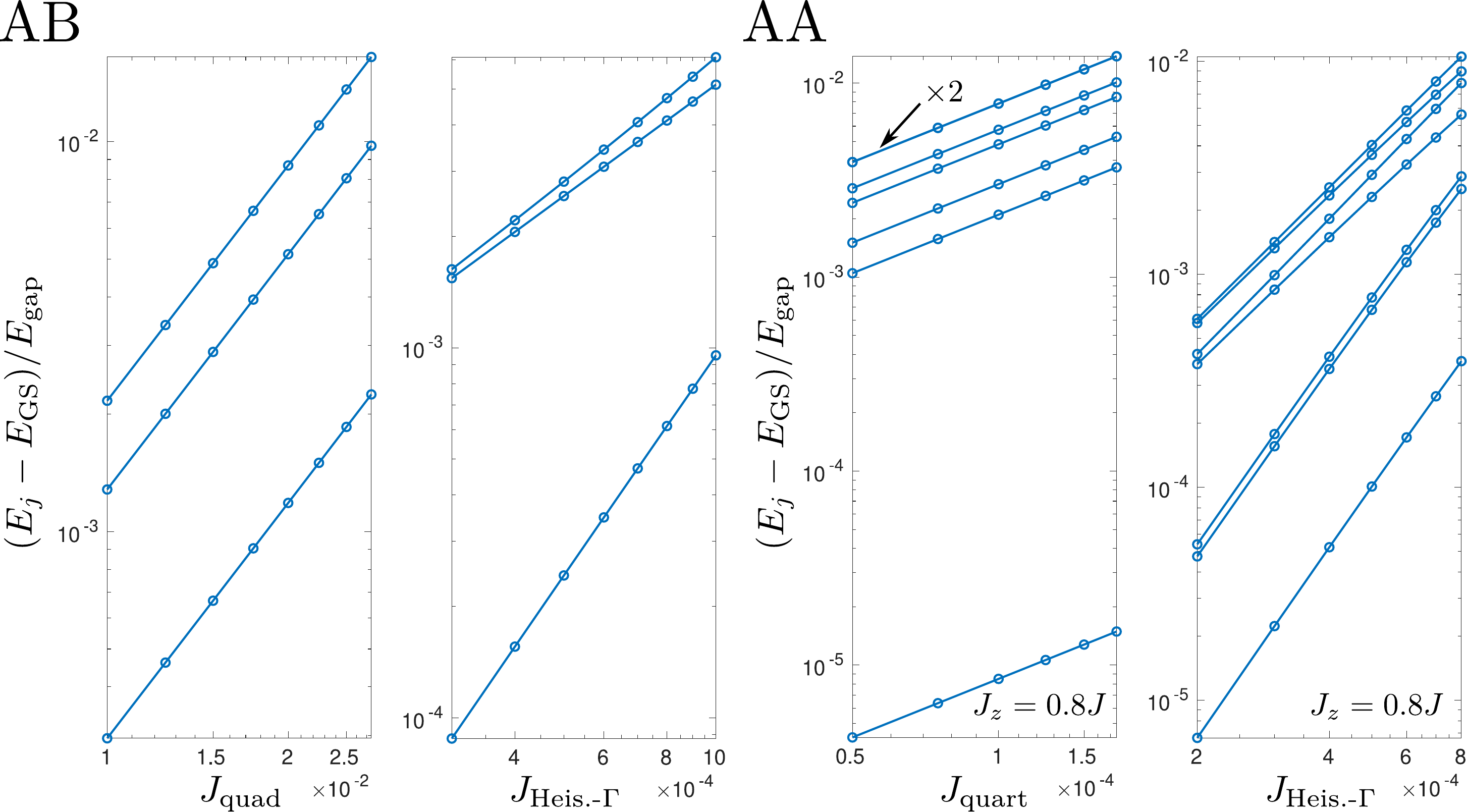}
   \caption{The energy differences $(E_j-E_{\text{GS}})/E_{\text{gap}}$ between the first 3 (or the first 7 in the AA case) excited states and the ground state as functions of different coupling strengths. $E_{\text{gap}}$ is the energy of the 4th (or the 8th in the AA case) excited state. In the AA case with quartic interactions, the 6th and 7th excited states have close energies indistinguishable in the plot. The energy difference changes as a power law with a fixed power for flux-preserving interactions. When interactions mix flux sectors, different energy levels may have different scaling dependences.}
   \label{fig: 2_vacs_energy_lift_parameters_app}
 \end{figure}
Anisotropic Kitaev couplings $(J_z \neq J_x,J_y)$ are important for the AA case, where the positions of the vacancies that preserve global spatial symmetries partially protect degeneracies.

Our results are summarized in Fig.~\ref{fig: 2_vacs_energy_lift_parameters_app}. They confirm our expectations that the ground state of generic KSL with vacancies is unique. We observed similar behavior on a 28-site cluster where two vacancies can be fully separated. In that case, only the AA case with quartic terms requires anisotropy for full degeneracy lifting.

\section{Comparison to other integrable models}
In this section, we briefly discuss vacancies in other two exactly solvable models \cite{Yao2007,Yao2011}. They both are defined on a two-dimensional decorated honeycomb lattice. To create a vacancy, we remove a triangle with three spins, similarly to \refcite{Sanyal2021}.
\subsection{Yao-Kivelson model }

This model has the following Hamiltonian
\begin{equation}
    H_{\text{YK}} = J_{\text{t}} \sum_{\langle \vect r \vect r' \rangle^\alpha \in \text{tri}} \sigma^\alpha_{\vect r} \sigma^\alpha_{\vect r'} + J_{\text{h}} \sum_{\langle \vect r \vect r' \rangle^\alpha \in \text{hex}} \sigma^\alpha_{\vect r} \sigma^\alpha_{\vect r'}
\end{equation}
with different coupling for triangles (tri) and hexagons (hex). The model is solved using Majorana fermions and fluxes (dodecagonal and triangular). The ground state occurs in the flux-free state, and the corresponding fermion spectrum is gapped for all finite values of the couplings except for the phase transition point $\boldsymbol{k}=(\pi,-\pi/\sqrt{3}),\quad J_{\text{h}}=\sqrt{3}_{\text{t}}$ and the limit $J_{\text{t}} \to \infty$. The two gapped phases are Abelian and non-Abelian.

In the Majorana representations with $\sigma^\alpha_{\vect r} = i \lambda_{\vect r} \xi^\alpha_{\vect r}$, the model reads
\begin{equation}
    H_{\text{YK}} = -iJ_{\text{t}}\sum_{\langle \vect r \vect r' \rangle^\alpha \in \text{tri}}\lambda_{\vect r} \hat{u}^\alpha_{\vect r \vect r'} \lambda_{\vect r'}-iJ_{\text{h}}\sum_{\langle \vect r \vect r' \rangle^\alpha \in \text{hex}} \lambda_{\vect r} \hat{u}^\alpha_{\vect r \vect r'} \lambda_{\vect r'}, 
    \label{eq: yao_kivel_mr}
\end{equation}
 The triangular and dodecagonal conserved fluxes are
\begin{equation}
\begin{aligned}
    W_{\text{tri}}&= \prod_{\langle \vect r \vect r' \rangle^\alpha \in \text{tri}} \hat{u}^\alpha_{\vect r \vect r'}=\prod_{ \vect r \in \text{tri}}\sigma^{\alpha_{\vect r}}_{\vect r},\\ W_{\text{dod}}&= \prod_{\langle \vect r \vect r' \rangle^\alpha \in \text{dod}} \hat{u}^\alpha_{\vect r \vect r'}=\prod_{ \vect r \in \text{dod}}\sigma^{\alpha_{\vect r}}_{\vect r},
\end{aligned}
\end{equation}
where $\alpha_{\vect r}$ corresponds to the third link that does not belong to the contour.
Removal of a triangle introduces new conserved quantities
\begin{equation}
    \tau^\alpha_{\vect R_\triangle} = \prod_{\vect r \in \text{dod}^\alpha \backslash \triangle} \sigma^{\alpha_{\vect r}}_{\vect r} \prod_{\vect r' \vect \in \triangle^\alpha} \sigma^{\alpha_{\vect r'}}_{\vect r'},
\end{equation}
where the first product excludes spins from missing triangle $\triangle$, and $\triangle^\alpha = \triangle +\hat e_\alpha$. This $\tau^\alpha_{\vect R_\triangle}$ operator has odd number of spins and it satisfies the Pauli algebra
\begin{equation}
        [\tau^\mu_{\boldsymbol{R}_\triangle} ,  \tau^\nu_{\boldsymbol{R}_\triangle}  ]= 2i \epsilon^{\mu\nu\kappa}\tau^\kappa_{\boldsymbol{R}_\triangle}  W_{\text{\vacflux}} W_{\triangle^x} W_{\triangle^y} W_{\triangle^z}~,
\end{equation}
where $W_{\text{\vacflux}}$ corresponds to a plaquette of three merged dodecagons.
In the Majorana representation, for fixed gauge choice, it reads
\begin{equation}
    \tau^\alpha_{\vect R_\triangle} = \frac{i}{2} \epsilon^{\alpha \beta \gamma} \xi^\beta_{\vect R_\triangle + \hat{e}_{\beta}} \xi^\gamma_{\vect R_\triangle + \hat{e}_{\gamma}}~,
\end{equation}
where $\xi^\alpha_{\vect R_\triangle + \hat e_\alpha}$ are dangling zero modes. Vectors $\hat e_\alpha$ connect neighbouring triangles along $\alpha$ direction, and we further use shorter notation $\vect R_\triangle + \hat{e}_{\alpha} = \vect R^\alpha_\triangle$.

As for the matter fermions, the vacancy introduces at least one zero mode. This can be understood from the following. For fixed $u^\alpha_{\vect r \vect r'}$, the Hamiltonian \eqref{eq: yao_kivel_mr} always has particle-hole symmetry: in the Majorana basis, $\mathcal{P}$ is the complex conjugation operator that satisfies $\mathcal{P} H_{\text{YK}} \mathcal{P}^{-1} = -H_{\text{YK}}$. Due to this symmetry, each $\epsilon_k$ state has a partner with energy $-\epsilon_k$, and the total number of non-zero states must be even.  In case of a vacancy, three Hamiltonian terms are discarded, and the total number of interacting Majorana operators is odd, so there is at least one Majorana zero mode. We numerically diagonalize the free-fermion Hamiltonian matrix and obtain only one zero mode for various coupling $J_{\text{t}},J_{\text{h}}$. Then, the zero energy subspace constraint has a form
\begin{equation}
    \psi_0 \xi^x_{\boldsymbol{R}^x_{\triangle}} \xi^y_{\boldsymbol{R}^y_{\triangle}} \xi^z_{\boldsymbol{R}^z_{\triangle}} = \pm 1~.
\end{equation}
Similarly to the Kitaev model, the magnetic moment reads
\begin{equation}
    \langle \sigma^\alpha_{\vect R^\alpha_\triangle} \rangle = \mathcal{N}_0 \langle \tau^\alpha_{\vect R_\triangle} \rangle~,
\end{equation}
where $\mathcal{N}_0$ is finite for phases with a spectral gap. For a gapless case, we can add the following interaction 
\begin{equation} 
\delta H = g \tau^\alpha_{\vect R_\triangle} \sigma^\alpha_{\vect R_\triangle^\alpha}, \quad [\tau^\alpha_{\vect R_\triangle},\delta H] = 0~,
\end{equation}
which gives finite magnetic moment.

Additionally, the spontaneous time-reversal symmetry breaking permits perturbation that couples $\tau^\alpha_{\vect R_\triangle}$ to a plaquette fluxes while preserving the time-reversal symmetry 
\begin{equation}
    H = H_{\text{YK}} + J_{\text{quad}}\tau^\alpha_{\vect R_\triangle} W_{\triangle^\alpha}~.
\end{equation}
In this case, the magnetic moment acquires polarization determined by $W_{\triangle^\alpha}$. This interaction is sufficient to remove degeneracy associated with a single vacancy and thus prevent the extensive ground state degeneracy.
\subsection{Yao-Lee model }

This model has the following Hamiltonian
\begin{equation}
\begin{aligned}
    H_{\text{YL}}&=\frac{1}{4}\sum_{\langle \vect r \vect r' \rangle^\alpha \in \text{hex}}J_{\alpha}\left[T_{\vect r}^{\alpha}T_{\vect r'}^{\alpha}\right]\left[\boldsymbol{\sigma}_{\vect r}\cdot\boldsymbol{\sigma}_{\vect r'}\right]~,
\end{aligned}
\end{equation}
where $T_{\vect r},\sigma_{\vect r}$ are independent spin-$1/2$ operators. This Hamiltonian has time-reversal and SU(2) symmetry.
Using the following Majorana fermions representation
\begin{equation}
\begin{aligned}
    &\sigma_{\vect r}^{\alpha}T_{\vect r}^{\beta}=i \lambda_{\vect r}^{\alpha} \xi_{\vect r}^{\beta},\quad\sigma_{\vect r}^{\alpha}=-\frac{\epsilon^{\alpha\beta\gamma}}{2}i \lambda_{\vect r}^{\beta}\lambda_{\vect r}^{\gamma}\quad T_{\vect r}^{\alpha}=-\frac{\epsilon^{\alpha\beta\gamma}}{2}i\xi_{\vect r}^{\beta}\xi_{\vect r}^{\gamma}\\
    &D=-i\lambda_{\vect r}^{x}\lambda_{\vect r}^{y}\lambda_{\vect r}^{z}\xi_{\vect r}^{x}\xi_{\vect r}^{y}\xi_{\vect r}^{z}~,\quad \hat u_{\vect r \vect r'}=-i\xi_{\vect r}^{\lambda}\xi_{\vect r'}^{\lambda}~,
\end{aligned}
\end{equation}
the Hamiltonian reads
\begin{equation}
\begin{aligned}
H_{\text{YL}}=\sum_{\langle \vect r \vect r'\rangle \in \text{hex} }  \frac{\hat u_{\vect r \vect r'}}{4}\left[J_x i\lambda_{\vect r}^{x}\lambda_{\vect r'}^{x}+J_y i\lambda_{\vect r}^{y}\lambda_{\vect r'}^{y}+ J_z i\lambda_{\vect r}^{z}\lambda_{\vect r'}^{z}\right]~.
\end{aligned}
\end{equation}
 The model has an extensive number of conserved fluxes 
\begin{equation}
    W_{\text{hex}} = \prod_{\langle \vect r \rangle \in \text{hex}} T^{\alpha_{\vect r}}_{\vect r}= \prod_{\langle \vect r \vect r' \rangle^\alpha} \hat u^\alpha_{\vect r \vect r'} ~,
\end{equation}
which include only $T^\alpha_{\vect r}$ operators. After fixing $W_{\text{hex}}$ values, the Hamiltonian becomes quadratic and has three species of fermions. Each species' excitations spectrum has two Majorana cones.
When introducing a vacancy, the local conserved quantities are
\begin{equation}
\begin{aligned}
    \tau^\alpha_{\boldsymbol{R}_\triangle} &= W^\alpha_{\boldsymbol{R}_\triangle} T^\alpha_{\boldsymbol{R}^\alpha_\triangle}~, 
\end{aligned}
\end{equation}
which satisfy the Pauli algebra
\begin{equation}
    [\tau^\mu_{\boldsymbol{R}_\triangle} ,  \tau^\nu_{\boldsymbol{R}_\triangle}  ]= 2i \epsilon^{\mu\nu\kappa}\tau^\kappa_{\boldsymbol{R}_\triangle}  W_{\text{\vacflux}}~.
\end{equation}
Thus, $\tau^\alpha_{\boldsymbol{R}_\triangle}$ provides locally encoded two-fold degeneracy of eigenstates, and, similarly to the Kitaev model, $N_{\text{v}}$ vacancies generate extensive degeneracy of the ground state $g \geq 2^{N_{\text{v}}}$.

In Majorana representation, creating a vacancy results into one zero mode for each species and three unpaired $\hat u_{\vect r \vect r'}$ operators, which gives six zero modes in total. The parity constraint projected to the zero mode subspace becomes $\psi^x_0 \psi^y_0 \psi^z_0 \xi^x_{\boldsymbol{R}^x_\triangle} \xi^y_{\boldsymbol{R}^y_\triangle} \xi^z_{\boldsymbol{R}^z_\triangle} = \pm i$, making the zero energy Hilbert space four-dimensional. One part of the degenerate subspace is spanned by 
\begin{equation}
\begin{aligned}
    \tau^\alpha_{\vect R_\triangle} & = \frac{i}{2} \vect \xi_{\vect R_\triangle} \times \vect \xi_{\vect R_\triangle}~, \\
    \vect \xi_{\vect R_\triangle} &= (\xi^x_{\vect R^x_\triangle},\xi^y_{\vect R^y_\triangle},\xi^z_{\vect R^z_\triangle})~.
\end{aligned}
\end{equation}
The remaining matter zero modes encode another Bloch sphere associated with the following operator
\begin{equation}
\begin{aligned}
    \vect s &= \frac{i}{2} \vect \psi_0 \times \vect \psi_0~.\\
\end{aligned}
\end{equation}
However, there is no local operator that would represent $\vect s$ in terms of $T,\sigma$ spins. 

In this model with two different type of spins on each site, it makes sense to consider separately $T$ and $\sigma$ vacancy moments. In the ground state, they are
\begin{equation}
\begin{aligned}
    \langle T^\alpha_{\vect R^\alpha_\triangle} \rangle &= \frac{i}{2} \epsilon^{\alpha \beta \gamma} \langle \xi^\beta_{\vect R^\alpha_\triangle} \xi^\gamma_{\vect R^\alpha_\triangle} \rangle\\
    &\propto \langle \xi^\alpha_{\vect R_{\triangle}^\alpha} \lambda^x_{\vect R_{\triangle}^\alpha} \lambda^y_{\vect R_{\triangle}^\alpha} \lambda^z_{\vect R_{\triangle}^\alpha} \rangle\\
    &\propto {\cal N}_0^3 \langle \xi^\alpha_{\vect R_{\triangle}^\alpha} \psi^x_{ 0} \psi^y_{0} \psi^z_{0} \rangle \propto {\cal N}_0^3 \langle \tau^\alpha_{\vect R_{\triangle}} \rangle~,\\
    \langle\sigma_{\boldsymbol{R}^\alpha_{\triangle}}^{\alpha}\rangle &=\frac{i}{2} \epsilon^{\alpha \beta \gamma} \langle \lambda^\beta_{\boldsymbol{R}^\alpha_{\triangle}}\lambda^\gamma_{\boldsymbol{R}^\gamma_{\triangle}} \rangle \\
    &= \frac{i}{2} \epsilon^{\alpha \beta \gamma} {\cal N}_0^2 \langle i\psi^\beta_{0}\psi^\gamma_{0} \rangle \propto {\cal N}_0^2 \langle s^\alpha \rangle~,
\end{aligned}
\end{equation}
where ${\cal N}_0 \to 0$ for a gapless spectrum in a thermodynamic limit. 
The finite contribution to the $T$-moment appears with the perturbation from the main text
\begin{align}
    \delta H_{\tau T} &= g_1 \tau^\alpha_{\vect R_\triangle} T^\alpha_{\vect R^\alpha_\triangle}~,\quad [\delta H_{\tau T},\tau^\alpha_{\vect R_\triangle}]=0~.
\end{align}
The finite $\sigma$-moment cannot be induced by similar interaction because it does not have a local analog of $\tau$. In a generic case, we can also think of interactions that couple $T$ and $\sigma$ spins. To construct the $\sigma$ part of the interaction, we use SU(2) scalars
\begin{align}
    \delta H^{BB}_{TT \sigma \sigma}& = g^{BB}_2 \sum_{\alpha} \epsilon^{\alpha \beta \gamma} T^\beta_{\vect R_\triangle^\beta} T^\gamma_{\vect R_\triangle^\gamma} \vect \sigma_{\vect R_\triangle^\beta} \cdot \vect \sigma_{\vect R_\triangle^\gamma}\\
    &\propto \vect \tau^B_{\vect R_{\triangle}} \cdot \vect K^B_{\sigma \sigma}\\
    &\propto \sum_{\alpha,\delta} \epsilon^{\alpha \beta \gamma} \xi^\beta_{\vect R_\triangle^\beta} \xi^\gamma_{\vect R_\triangle^\gamma} \lambda^\delta_{\vect R_\triangle^\beta} \lambda^\delta_{\vect R_\triangle^\gamma} \\
    \delta H^{BA}_{TT \sigma \sigma}& = g^{BA}_2 \sum_{\alpha} \prod_{\substack{\vect r \in \Gamma_{\vect R_\triangle^\beta \vect r^\beta} \\ \vect r' \in \Gamma_{\vect R_\triangle^\gamma \vect r^\gamma}}}  \epsilon^{\alpha \beta \gamma} T^{\alpha_r}_r  T^{\alpha_{r'}}_{r'} \vect \sigma_{\vect r^\beta} \cdot \vect \sigma_{\vect r^\gamma}  \\
        &\propto \vect \tau^B_{\vect R_{\triangle}} \cdot \vect K^A_{\sigma \sigma}\\
    &\propto \sum_{\alpha,\delta} \epsilon^{\alpha \beta \gamma} \xi^\beta_{\vect R_\triangle^\beta} \xi^\gamma_{\vect R_\triangle^\gamma} \lambda^\delta_{\vect r^\beta} \lambda^\delta_{\vect r^\gamma}~,
\end{align}
where a vacancy is at the $A$ sublattice, and the product of $T$ along $\Gamma_{\vect R_\triangle^\gamma \vect r^\gamma}$ is such that it couples one dangling fermion to $\vect \lambda_{\vect r}$. Operators $\vect K^{B(A)}_{\sigma \sigma}$ are SU(2) invariant and time-reversal odd due to a pair of $\sigma^\alpha_{\vect r}$ and product of an odd number of $T^\alpha_{\vect r}$. They commute with conserved fluxes. In Majorana representation, these terms couple the dangling fermions to a pair of $\vect \lambda_{\vect r}$ from the same or different sublattices. That permits standard RKKY interaction between $\tau^\alpha_{\triangle}$.  
 Interestingly, the quadratic coupling from the main text that could induce a non-zero $\langle \sigma^\alpha_{\vect r} \rangle$ and non-typical RKKY interaction violates the SU(2) symmetry.
 \section{Effective couplings for the Kitaev-Heisenberg-$\Gamma\text{-}\Gamma'$ model}
\begin{table}[ht]
    \centering
    \renewcommand{\arraystretch}{1.8} 
    \begin{tabular}{cccccc}
        \toprule
        \textbf{Coupling} & \multicolumn{2}{c}{\textbf{Only $\text{Heis.-}\Gamma$}} & \multicolumn{2}{c}{\textbf{Including $\Gamma'$}} \\ 
        \midrule
        $g$  
        & $\frac{J_{\Gamma}^4}{J^3}~$ 
        & \raisebox{-1.5em}{\includegraphics[width=2cm]{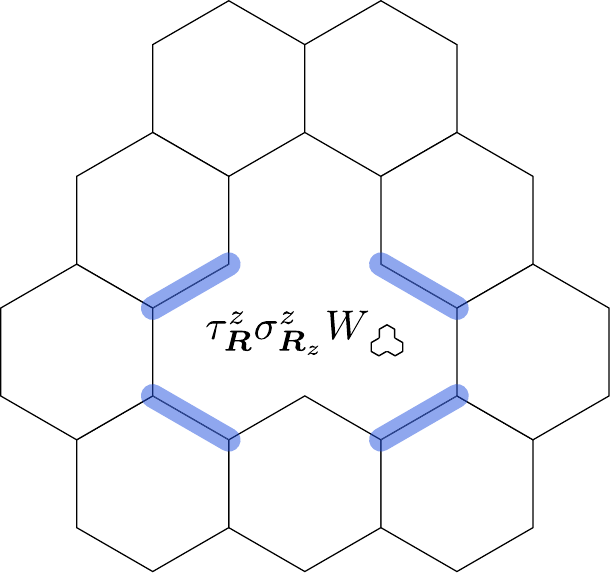}} 
        & $\frac{J_{\Gamma}^4}{J^3}$ 
        & \raisebox{-1.5em}{\includegraphics[width=2cm]{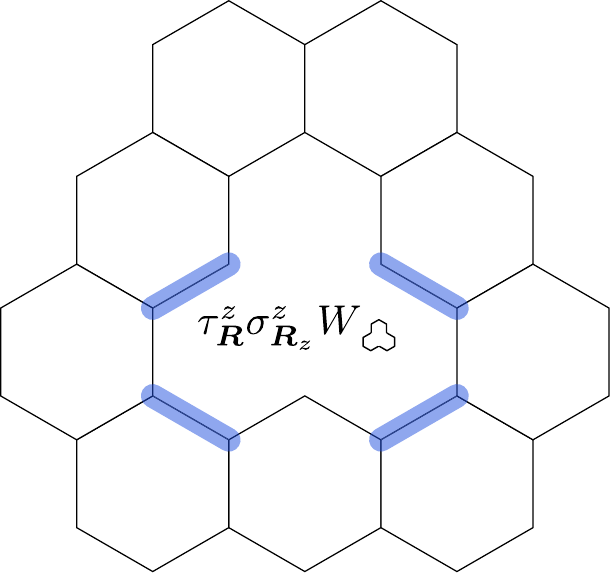}} 
         \\ 
          
        &  $\frac{J_{\text{Heis.}}^3 J_\Gamma^3}{J^5}$ 
        & \raisebox{-1.5em}{\includegraphics[width=2cm]{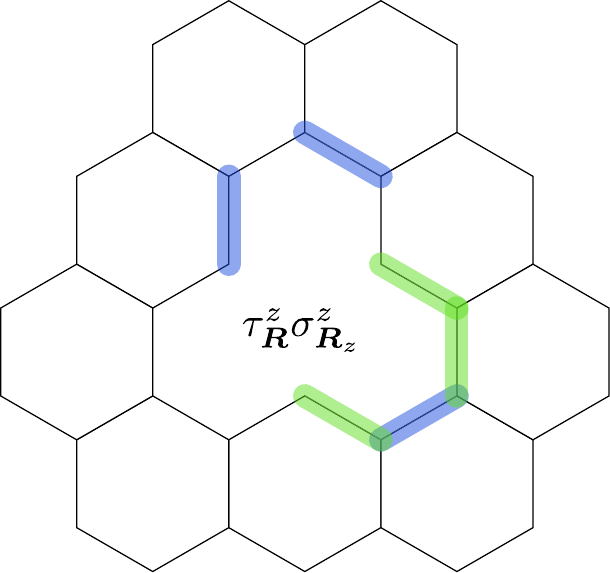}} 
        & $\frac{J_{\text{Heis.}}^3 J_\Gamma^2 J_{\Gamma'}}{J^5}$ 
        & \raisebox{-1.5em}{\includegraphics[width=2cm]{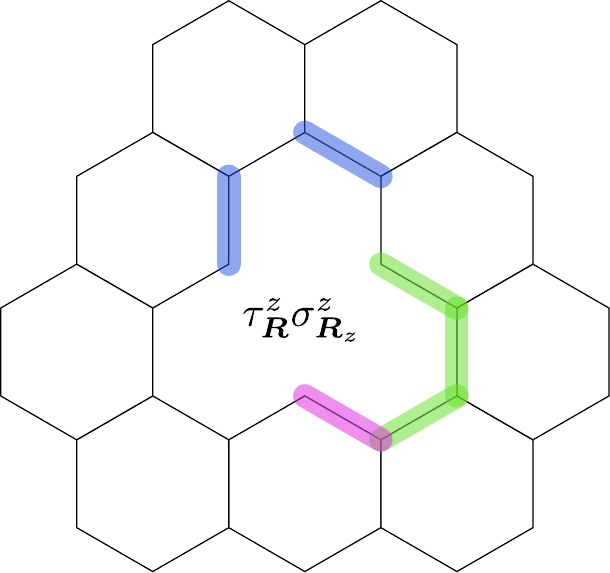}} 
         \\  
        $t$  
        & $\frac{J_{\text{Heis.}}^2 J_\Gamma^2}{J^3}$ 
        & \raisebox{-1.5em}{\includegraphics[width=2cm]{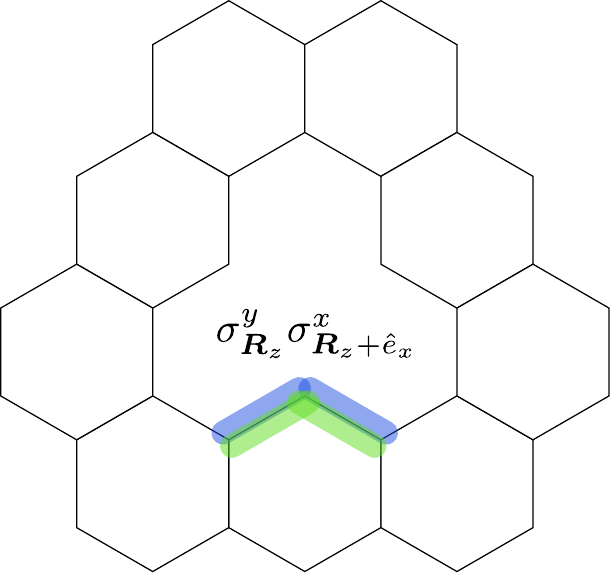}} 
        & $J_{\Gamma'}$ 
        & \raisebox{-1.5em}{\includegraphics[width=2cm]{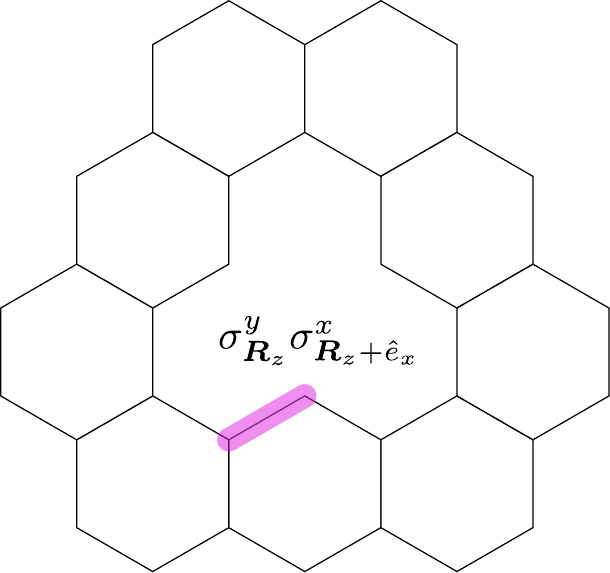}} 
         \\
        $f^\mu$  
        & $\frac{J^2_{\text{Heis.}} J_\Gamma^2}{J^3}$ 
        & \raisebox{-1.5em}{\includegraphics[width=2cm]{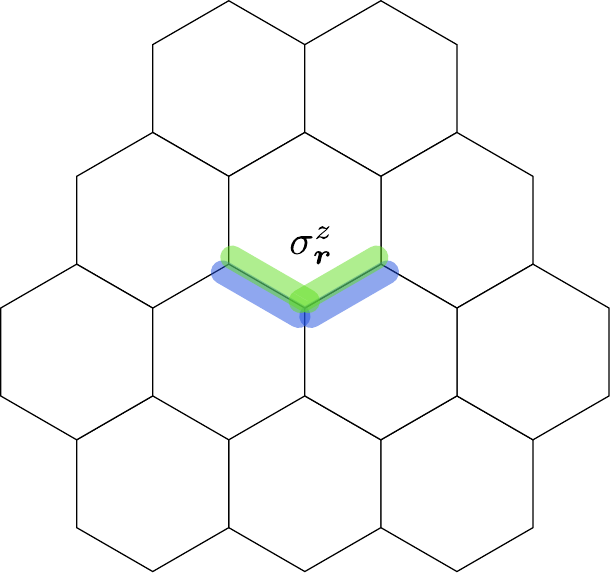}} 
        & $J_{\Gamma'}$ 
        & \raisebox{-1.5em}{\includegraphics[width=2cm]{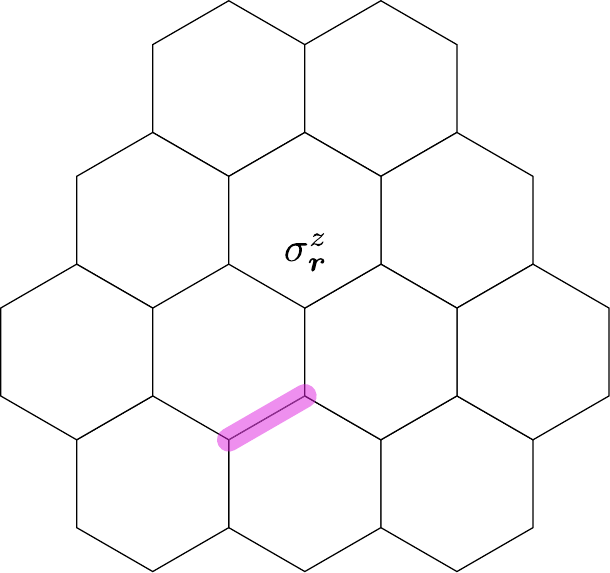}} 
         \\          
        \bottomrule
    \end{tabular}
    \caption{Scaling of the effective couplings as functions of $\text{Heisenberg-}\Gamma\text{-}\Gamma'$ interactions. The colored bonds represent examples of these interactions and are shown in green, blue, and magenta, respectively.}
\label{tab: couplings_scaling}
\end{table}
Most microscopic studies propose a Hamiltonian of the following form
\begin{equation}
\begin{aligned}
 H_{\text{KH} \Gamma \Gamma'} &= J \sum_{  \vect r,\mu }  \sigma^{\mu}_{\vect r} \sigma^{\mu}_{\vect r+\hat e_\mu} \\ & + J_{\text{Heis.}}  \sum_{\langle\vect  r \vect r'\rangle} \vect \sigma_{\vect r} \cdot \vect \sigma_{\vect r'}\\ &+J_{\Gamma} \sum_{\vect r,\mu}|\epsilon_{\mu\nu\kappa}|\sigma_{\vect r}^{\nu} \sigma_{\vect r+\hat e_\mu}^{\kappa}\\ &+J_{\Gamma'} \sum_{\vect r,\mu}|\epsilon_{\mu\nu\kappa}|\left( \sigma_{\vect r}^{\nu} \sigma_{\vect r+\hat e_\mu}^{\mu} + \sigma_{\vect r}^{\mu} \sigma_{\vect r+\hat e_\mu}^{\nu} \right)~,
 \label{eq: full_hamilt}
\end{aligned}
\end{equation}
known as the Kitaev-Heisenberg-$\Gamma$-$\Gamma '$ model. In our main text, we instead use the effective couplings
\begin{equation}
\begin{aligned}
    H_{ \sigma \tau} &= g\sum\nolimits_\mu  \tau^\mu_{\vect R} \sigma^\mu_{\boldsymbol{R}+\hat e_\mu}~,\\ \quad H_{\psi\lambda} &=i t \sum\nolimits_{\mu}\psi_{\vect R}^\mu \lambda_{\vect R_{\mu}}~,\\ \quad     \delta\sigma_{\vect r}^\mu &\sim i f^\mu_{\vect r\vect r'\vect r''}  \lambda_{\vect r'}\lambda_{\vect r''}~,
    \label{eq: eff_couplings}
\end{aligned}
\end{equation}
which allow us to make predictions about the vacancy magnetization and interactions between $\tau$ spins. 
To relate the effective description to the microscopic one, we determine the dependence of the parameters $g,t,f$ on the microscopic couplings $J_\text{Heis}, J_{\Gamma}, J_{\Gamma'}$ in the asymptotic regime where all these couplings are much smaller than $J$. We note that numerics find the Kitaev phase to be unstable beyond this limit.

 We determine the scaling of effective couplings by considering the products of terms from Eq. \ref{eq: full_hamilt} that, in the perturbation theory, result in Eq. \ref{eq: eff_couplings}, similarly to \refcite{Song2016}. Our results are summarized in Table \ref{tab: couplings_scaling}. We obtain that effective interactions similar to $ \tau^\alpha_{\vect R} \sigma^\alpha_{\vect R_\alpha}$ appear in the fourth order in $J_{\Gamma}$. In particular,
\begin{equation}
H_{ \sigma \tau} = g \tau^\alpha_{\vect R} \sigma^\alpha_{\vect R_\alpha} W_{\text{\vacflux}} = \pm \frac{g}{2} \epsilon_{\mu \nu \alpha} \psi^\mu_{\vect R} \psi^\nu_{\vect R}
 \psi^\alpha_{\vect R}
 \lambda_{\vect R_\alpha}~, 
\end{equation}
where $\vect R_\alpha = \vect R + \hat e_\alpha$, 
is the eight-spin operator that can be constructed from the product of four $\Gamma$ terms. On the contrary, the six-spin operator from the main text arises in the third order in Heisenberg and the third order in $\Gamma$ interactions. 
Interestingly, the presence of the interaction $\Gamma'$ explicitly includes first-neighbor $t$ hoppings and generates $f^\mu$ in the first order in $J_{\Gamma'}$, which can promote more pronounced magnetization and stronger `phantom' spin interactions.
\end{document}